\documentclass[letterpaper,twocolumn,10pt]{article}
\usepackage{usenix-2020-09}

\usepackage{tikz}
\usepackage{amsmath}
\usepackage{subfig}
\usepackage{balance}
\usepackage{filecontents}
\usepackage{paralist}
\usepackage{booktabs}
\usepackage{amssymb}

\newcommand{\new}[1]{\textcolor{black}{#1}}
\newcommand{\review}[1]{\textcolor{black}{#1}}

\begin{document}

\date{}

\title{Hand Me Your PIN! \\
Inferring ATM PINs of Users Typing with a Covered Hand}

\author{
{\rm Matteo Cardaioli}\\
University of Padua, Italy\\
GFT Italia, Italy

\and
{\rm Stefano Cecconello}\\
University of Padua,\\Italy

\and
{\rm Mauro Conti}\\
University of Padua,\\Italy

\and
{\rm Simone Milani}\\
University of Padua,\\Italy

\and
{\rm Stjepan Picek}\\
Delft University of Technology and \\ Radboud University, The Netherlands

\and
{\rm Eugen Saraci}\\
University of Padua,\\Italy
} 

\maketitle

\pagestyle{empty}
\begin{abstract}
Automated Teller Machines (ATMs) represent the most used system for withdrawing cash. The European Central Bank reported more than 11 billion cash withdrawals and loading/unloading transactions on the European ATMs in 2019.
Although ATMs have undergone various technological evolutions, Personal Identification Numbers (PINs) are still the most common authentication method for these devices.
Unfortunately, the PIN mechanism is vulnerable to shoulder-surfing attacks performed via hidden cameras installed near the ATM to catch the PIN pad. 
To overcome this problem, people get used to covering the typing hand with the other hand. While such users probably believe this behavior is safe enough to protect against mentioned attacks, there is no clear assessment of this countermeasure in the scientific literature.

This paper proposes a novel attack to reconstruct PINs entered by victims covering the typing hand with the other hand. 
We consider the setting where the attacker can access an ATM PIN pad of the same brand/model as the target one.
Afterward, the attacker uses that model to infer the digits pressed by the victim while entering the PIN. 
Our attack owes its success to a carefully selected deep learning architecture that can infer the PIN from the typing hand position and movements. 
We run a detailed experimental analysis including 58 users. With our approach, we can guess 30\% of the 5-digit PINs within three attempts -- the ones usually allowed by ATM before blocking the card.
We also conducted a survey with 78 users that managed to reach an accuracy of only 7.92\% on average for the same setting.
Finally, we evaluate a shielding countermeasure that proved to be rather inefficient unless the whole keypad is shielded.



\end{abstract}

\section{Introduction}
\label{sec:introduction}

The wide deployment of various Cyber-Physical Systems (CPS) has a significant impact on our daily lives.
Unfortunately, the increased use of CPS also brings more threats to users. This is especially pronounced considering new attack vectors that use machine learning approaches.
As such, threats become a global issue, and the need to design secure and robust systems increases.
One common security mechanism in devices like Automated Teller Machines (ATMs) and Point of Sale (PoS) depends on the security provided by the Personal Identification Numbers (PINs).
While ATMs and PoS devices are widely used~\footnote{https://sdw.ecb.europa.eu/reports.do?node=1000001407}, 
many people do not consider security risks and defenses beyond those commonly mentioned~\footnote{https://www.hsbc.com.hk/help/cybersecurity-and-fraud/atm-scams/}: i) hide the PIN while typing, and i) make sure no one watches the screen (shoulder-surfing attack).
In the context of financial services, ISO 9564-1~\cite{9564} specifies the basic security principles for PINs and PIN entry devices (e.g., PIN pads). \new{For example, to mitigate the shoulder surfing attacks~\cite{eiband2017understanding,binbeshr2020systematic}, the standard indicates that i) PIN digits must not be displayed on a screen, and ii) the duration and type of feedback sound emitted must be the same for each key.}
Consequently, as a compromise between security and usability, PIN entry systems display a fixed symbol (e.g., a dot) to represent a digit being pressed and provide the same audio feedback (i.e., same tone, same duration) for all keys. 
Thus, the combination of security mechanisms enforced by standards and the common precaution measures taken by users should provide sufficient protection. 
Unfortunately, the attackers also improve their approaches over time and consider more sophisticated attacks.


The security of ATM and PoS devices is of great concern as millions of such devices are used~\cite{del2020mobile}. Resourceful attackers that succeed in attacking even a small percentage of those devices can cause significant damage considering costs and public perception.
This problem is especially pronounced as last years brought significant developments in the attack techniques~\cite{balagani2019pilot, cardaioli2020your, liu2019human}.
At the same time, attacking ATM or PoS devices is not easy, especially if considering realistic settings. Most of the state-of-the-art attacks can be defeated by a careful user covering the PIN that is entered. Recent results that consider thermal cameras are also difficult to succeed, depending on the keypad type and the time users spend operating the device. The attacker can also use timing or acoustic attacks to infer information about the entered digits, but they are not as effective as the state-of-the-art attacks since they require additional information such as thermal residues~\cite{cardaioli2020your}, making it challenging to apply realistically such attacks.

This work proposes a novel attack aiming to reconstruct PINs entered by victims that cover the typing hand by the other hand. More precisely, we leverage the advances in the deep learning domain to develop an attack predicting what PIN is entered based on the position of the user's hand and the movements while pressing the keys. Our attack gives high accuracy rates even in the cases when the user perfectly covers the typing hand. What is more, our attack reaches higher accuracy values than previous works that needed to consider several sources of the information at the same time (timing, sound, and thermal signatures)~\cite{cardaioli2020your}.

Our attack considers a profiling setting where the attacker has access to a PIN pad that is identical (or at least similar) to the one used by the victim. Then, we build a profiling model that can predict what digit is entered on the target device.
This is the first attack on PIN mechanisms that works even when the PIN is covered while being entered to the best of our knowledge.
Our attack demonstrates that the ATM and PoS security mechanisms are insufficient, and we must provide novel defenses to mitigate attackers.
We made our code and datasets publicly available at~\url{https://spritz.math.unipd.it/projects/HandMeYourPIN}.

\textbf{Main contributions} 
\begin{compactitem}
\item We propose a novel attack to infer PINs from videos of users covering the typing hand with their non-typing hand.
\item We demonstrate that our attack can reconstruct 30\% of 5-digit PINs and 41\% of 4-digit PINs within three attempts, showing that hiding the PIN while typing is insufficient to ensure proper protection.
\item We evaluate our attack via extensive experiments, collecting videos of 5\,800 5-digit PINs entered in a simulated ATM by 58 participants. We conduct a study to assess humans' accuracy in inferring covered PINs from videos. We show that our attack outperforms humans, achieving a four-fold improvement on reconstructing 5-digits PINs within three attempts.
\item We pre-process our dataset, and we make it publicly available to the research community. We hope this is beneficial to understand the problem better and propose possible solutions.
\item We discuss several countermeasures that would make the attack more difficult to conduct. We perform an analysis on the attack performance when covering the PIN pad (coverage 25\%, 50\%, 75\%, and 100\%) and show that attacks are possible even when using this countermeasure.
\end{compactitem}



\section{Threat Model}
\label{sec:ThreatModel}

The attack is performed when a victim interacts with a generic ATM keypad and types the PIN. The ATM is equipped with a PIN pad that emits a feedback sound when a key is pressed. The feedback sound is the same for all the keys of the PIN pad. The ATM is equipped with a monitor where obfuscated symbols appear when users enter a PIN to mask the entered digits.
We do not assume that the ATM or its PIN pad have been compromised during the attack. 
\review{
Our approach can be considered an alternative to card-skimmer attacks since we consider a different source of information to retrieve the PIN. Usually, card-skimming attacks rely on fake PIN pads that directly record the entered digits~\cite{scaife2018fear}, while our approach infers the PINs from a video.}

\subsection{Attacker}
\label{sec:Attacker}

The attacker is a malicious user aiming to steal the victim's secret PIN. 
\new{The attacker can place a hidden camera near the ATM to record the PIN pad. We make no assumptions about the type of camera used by the attacker except that it records in the visible spectrum~\footnote{We will use cheap and easily-concealable video sensing equipment, where standard RGB cameras fit such requirements.}. We assume that the camera can easily be hidden close to the ATM while keeping a direct view of the PIN pad (i.e., a pinhole camera if the attacker has access to the ATM~\footnote{\url{https://www.sperrywest.com/cameras/}} or any standard camera placed outside the ATM chassis).}
We also do not assume any specific position for the camera, but we discuss various camera placements' advantages.
We primarily consider the scenario where the attacker uses only one camera, but we also discuss the attack performance when using multiple cameras.
 
\new{The attack may take place together with different card stealing approaches: i) card skimming both on chip~\cite{bond2014chip} or magnetic stripe~\cite{scaife2018fear} (currently, the two payment-enabling technologies work together~\cite{emvMagneticStrip}), ii) exploiting a relay attack on a contactless card~\cite{gupta2020survey}, and iii) physically stealing the victim's card.}

We assume a profiling side-channel attack where the side-channel information comes from the video of the victim's hand while entering the PIN.
More precisely, side-channel information is the position of the victim's hand and the hand movements (both moving the hand/fingers to reach different keypads or movements observable due to muscle movements while a certain keypad is pressed).
The attacker can record a number of PINs entered on a copy of the ATM device and train a profiling model to predict what key is pressed. 
The attacker can retrieve the timestamps when the victim has typed the single keys on the keypad and can do so by listening to the audio of the video recording. There are two different types of sound clues that the attacker can exploit: the first one is the feedback sound made by the keypad when a key is pressed~\cite{cardaioli2020your}, the second one is the sound of the physical button of the keypad that is pressed.
External noise does not prevent the attacker from extracting the keypresses, as the camera is close enough to the keypad. As such, the sound can still be identified in the audio track. If, for any reason, the attacker has no way to retrieve the timestamps from the recorded audio (or if there is no audio at all), it is possible to place the camera to record both the keypad and the screen of the ATM~\cite{balagani2019pilot}. This allows the attacker to extract the keypresses' timing by looking at the PIN masking symbols appearing on the screen. Common masking symbols are usually dots and asterisks. 
The attacker can use any method to build a profiling model to predict what keys are pressed.
We consider the top three predictions as a measure of success since most ATMs will allow entering the PIN three times before blocking the card.
Finally, we do not assume that the PIN has any specific structure (pattern) that could be used to improve the attack performance further.

\subsection{Victim}
\label{sec:Victim}
We assume that the victim adopts basic countermeasures against card-skimming attacks, such as covering the hand while entering the PIN. The attacker does not need to be there when the victim types the PIN, as the attacker can freely access the camera's recorded video, either remotely or at a different time.


\section{Attack Approach}
\label{sec:attack}
Our attack assumes that the attacker has access to a training device and controls the PIN selection.
Additionally, the attacker knows the layout of a target device and will select the training device to be similar.
The attacker does not know the specific person to be attacked or the PIN for the attacked device.

\subsection{Attack Phases}
\label{sec:AttackPhases}

\begin{figure}[!ht]
	\centering
	\includegraphics[width=0.6\linewidth]{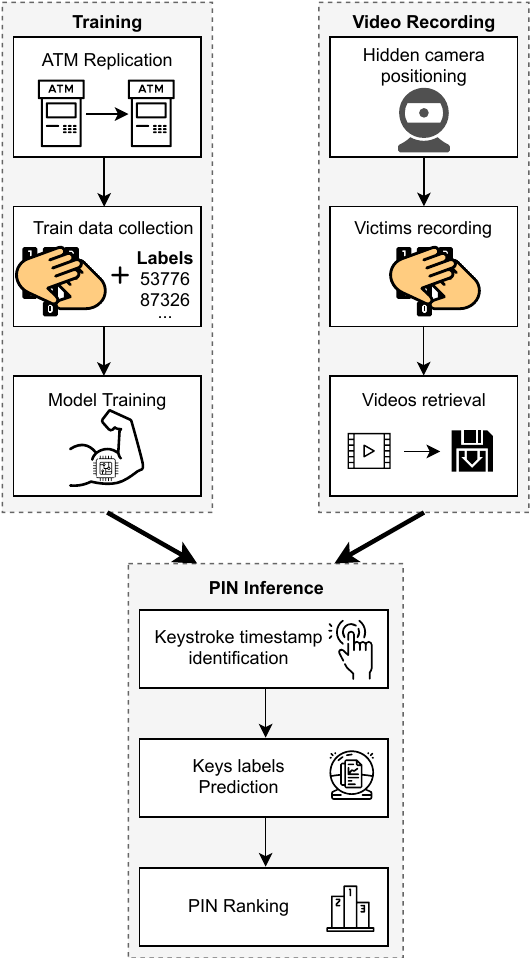}
	\caption{The attack step-by-step. The data collection process does not necessarily need to happen before the attacker steals the victim’s PIN. Still, it is a required step of the attack.}
	\label{fig:AttackStepByStep}
\end{figure}

We can divide the attack into three phases: Phase A -- Training, Phase B -- Video Recording, and Phase C -- PIN Inference. Figure~\ref{fig:AttackStepByStep} shows the required steps for the attack.




$ $\\

\textbf{Phase A -- Training}

\new{The attacker selects an ATM as the target of the attack. Next, the attacker sets up a replica of the target ATM. This replica does not have to be a faithful copy of the original, as our model takes in as input a crop around the keypad of the ATM. Therefore, the attacker must use a keypad similar to the one on the target ATM. The best situation is when the attacker can retrieve the same PIN pad model. Alternatively, the attacker can also use PIN pads that differ slightly (e.g., the key spacing can vary by a few millimeters). Note that the layout of ATM PIN pads has to follow the ISO 9564 standard~\cite{9564}.
}
The attacker uses the ATM replica to build the training set, simulating the victim's behavior while entering the PIN (i.e., covering the typing hand). The attacker must enter sequences of PINs on the replica PIN pad, including all ten digits (i.e., all the digits must be included in the training set).
Without losing generality, the attacker can use a USB PIN pad that logs the keys pressed and the corresponding timestamps. The attacker uses this information to segment the videos and labels them. Leveraging the logs, the attacker builds a training set containing, for each key pressed, a sequence of frames and the corresponding label (digits).
Finally, the attacker trains the predictive model on the collected training set. \new{For a detailed discussion on the implemented model, we refer readers to Section~\ref{sec:PredictionModels}}.\\

\textbf{Phase B -- Video Recording}

The attacker hides a camera near the target ATM to record the PIN pad. There are multiple places where the camera can be placed, and depending on this, the attack can be easier or more challenging to succeed.
The camera records the victim while entering the PIN and covering the PIN pad with the non-typing hand. The attacker retrieves the recorded video from the remote camera.\\

\textbf{Phase C -- PIN Inference}

The attacker's goal is to infer the victim's PIN based on the video recorded during the PIN entering.
First, the attacker retrieves the timestamps from the recorded video. The attacker can use both the pressed keys' feedback sound or the masking symbols appearing on the screen while the victim enters the PIN to perform this task.
Leveraging the timestamps, the attacker performs the same procedure as in Phase A to generate an attack set. Differing from the training set, the attack set contains a sequence of frames for each victim key pressed but no information about the related label.
The adversary detects in the attack set the frames corresponding to a PIN entry, and splits the video into $N$ sub-sequences where $N$ represents the number of digits composing the PIN.
For each sub-sequence, the adversary applies the model trained in Phase A. The model provides the probability of each class (i.e., the ten possible digits) to be the one corresponding to the input sub-sequence.
Exploiting the $N$ sub-sequences predictions, the attacker builds a rank of PINs in the descending order of their probabilities. In particular, the probability of a PIN corresponds to the product of the predicted probabilities of its digits.

\subsection{Attack Settings}
\label{sec:AttackSettings}

We consider three realistic attack scenarios:
\begin{compactenum}
    \item \textbf{Single PIN pad} scenario: the attacker knows the model of the target PIN pad and obtains a copy of it to carry out the training phase. While this scenario may seem unrealistic, we note it is not difficult to obtain a specific keypad copy. Indeed, the attacker can easily obtain information about the ATM to be attacked and then buy the keypad with the same layout. Naturally, there can be certain differences concerning how sensitive the keypad is (for instance, due to usage, pads can become somewhat more difficult to press), but our experiments indicate such differences are not substantial enough to pose issues for deep learning models.

    \item \textbf{PIN pad independent} scenario: this is the most challenging scenario. The attacker does not know or cannot retrieve the model of the target PIN pad. The training phase is performed on a PIN pad with similar characteristics to the target (e.g., shape, distance between keys, keys layout, and the sensitivity of keys).
    
    \item \textbf{Mixed} scenario: as for the \textit{Single PIN} scenario, the attacker knows the target PIN pad model. In this case, the training is performed on two PIN pads: a copy of the target and at least one PIN pad with similar characteristics. Considering several keypads in the training set makes sense when 1) the attacker is not certain about the keypad model, 2) the attacker assumes that the keypad will behave differently due to environmental conditions, 3) the attacker aims to attack multiple types of keypads (ATMs) with the same machine learning model, and 4) for any reason, the attacker did not manage to obtain enough training examples with a single keypad. We also note that using more keypads in the training set makes the training process more difficult and reduces the chances to overfit (i.e., we can consider different keypads as one keypad with noise, having the regularization effect~\cite{10.1162/neco.1995.7.1.108}).
\end{compactenum}

\subsection{Camera Positions}
\label{sec:CameraPositions}

Since our threat model allows the arbitrary position of the camera, we discuss several representative scenarios.
We consider positions at the top of the ATM preferable for the attacker as lower positions of the camera result in no visibility of the hand pressing the keys if the other hand is covering it.
We also consider settings at the front side of the chassis as they give better visibility for the attacker and are significantly more difficult for the victim to notice the camera.

Then, without loss of generality, we can discuss three main positions for the camera to provide good results. The camera can be positioned in the top left, center, or right corner. 
If the camera is positioned in the right corner and the person entering the PIN is right-handed, it will be easier to observe the entered digits. The same happens for the camera in the left corner and the left-handed person.
However, if the camera is in the center position, it does not favor any specific setting, making it the most general setting, but it also makes it somewhat more challenging to conduct the attack than the left/right position and left/right-handed persons.
We will concentrate on the top center position of the camera mounted on the chassis's front side.

\section{Experimental Setting}
\label{sec:ExperimentalSetting}
To assess the feasibility of our attack on all the scenarios described in Section~\ref{sec:attack}, we collected two datasets containing videos of people covering their typing hands while entering PINs. This section first illustrates the differences between the considered PIN pads and then describes our data collection procedure. Finally, the adopted video pre-processing, the setup used to run the experiments, and the implemented deep learning models are presented.

\subsection{Devices under Test}

\begin{figure}[htb]
\centering
\subfloat[\emph{DAVO LIN Model \textit{D-8201 F}}\label{fig:PINPad1}]
{\includegraphics[width=0.4\linewidth]{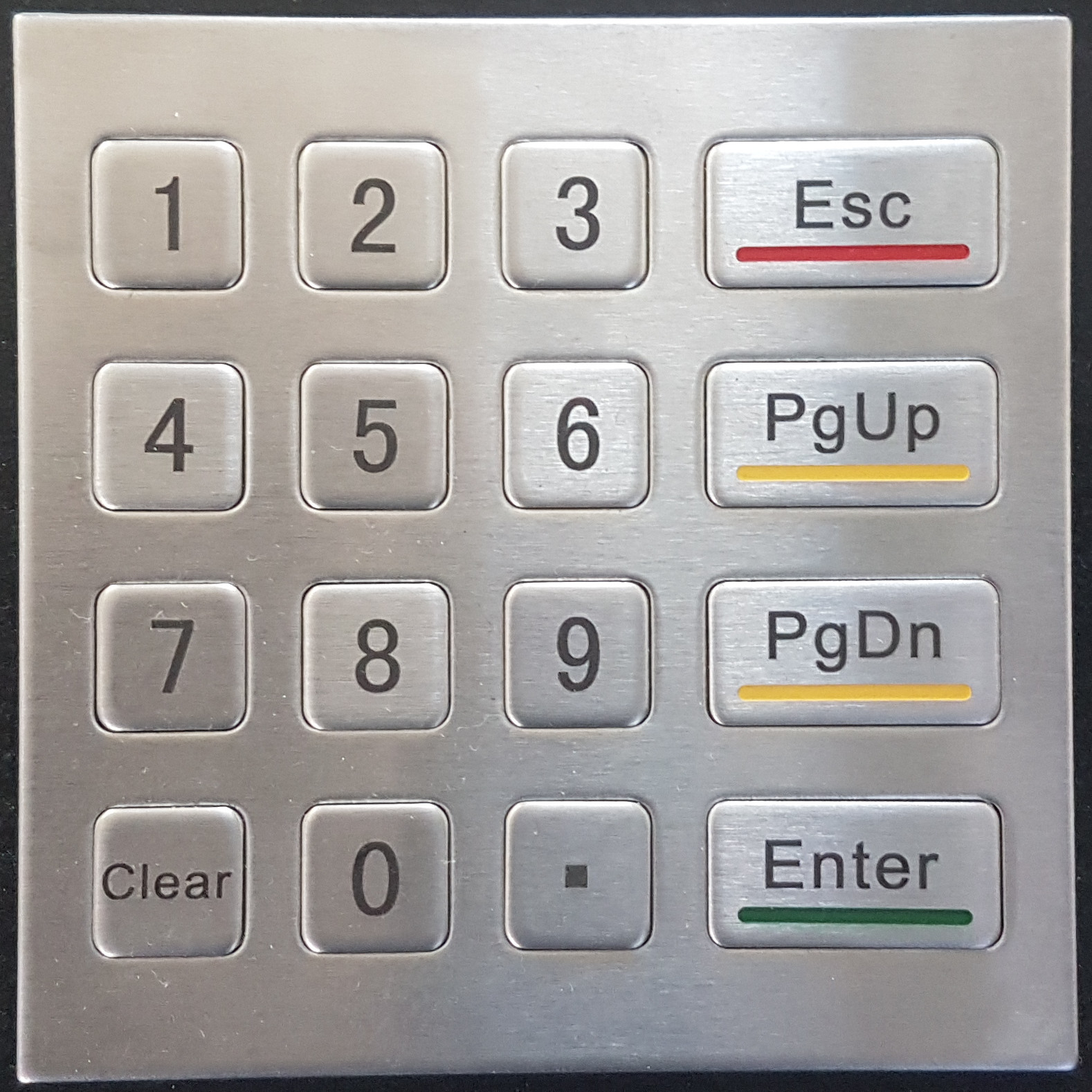}}\quad
\subfloat[\emph{DAVO LIN Model \textit{D-8203 B}}\label{fig:PINPad2}]
{\includegraphics[width=0.538\linewidth]{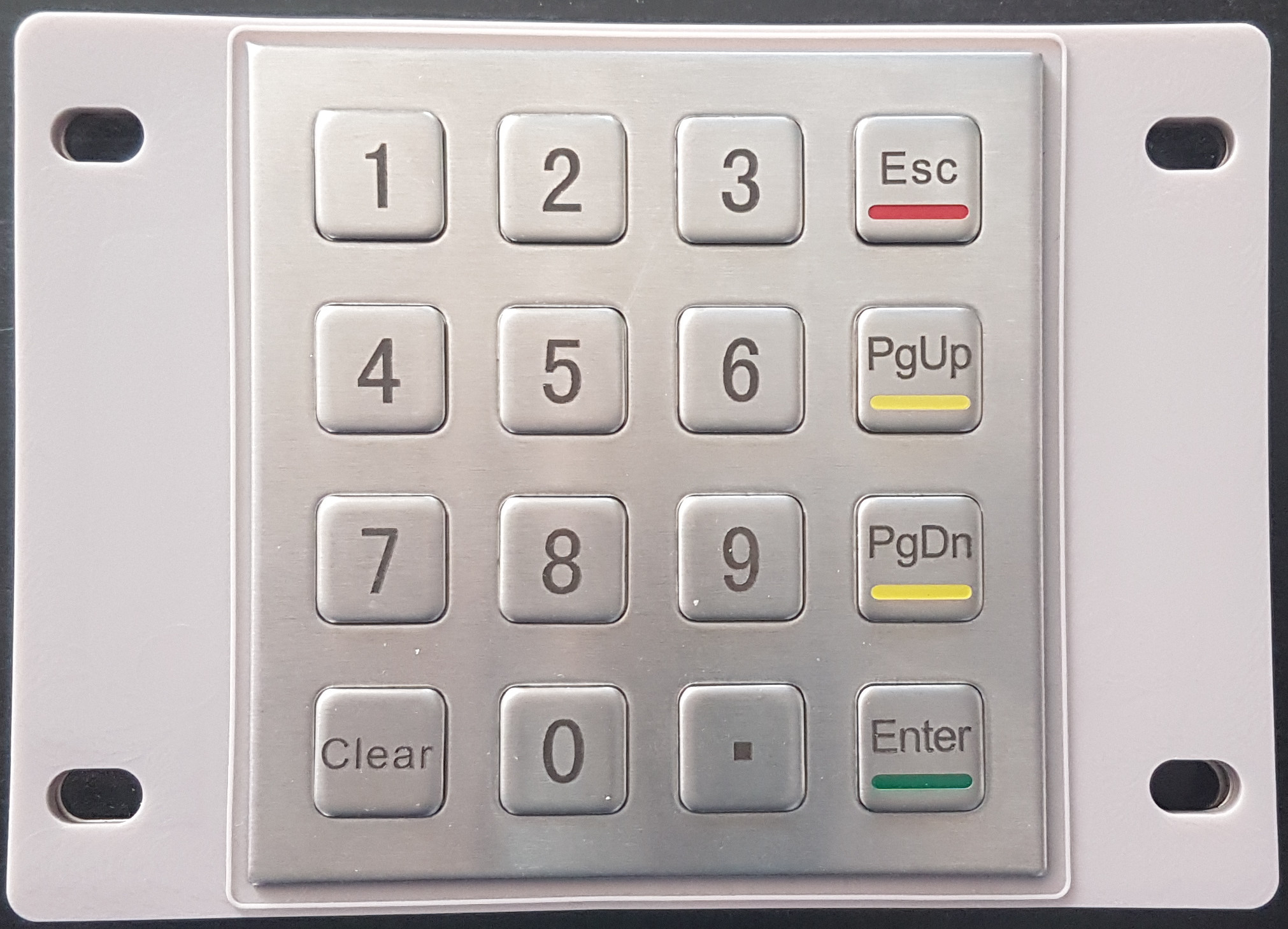}}\\
\caption{The PIN pads used in the data collection.}
\label{fig:PINPads}
\end{figure}

\begin{figure}[htb]
	\centering
	\includegraphics[width=0.65\linewidth]{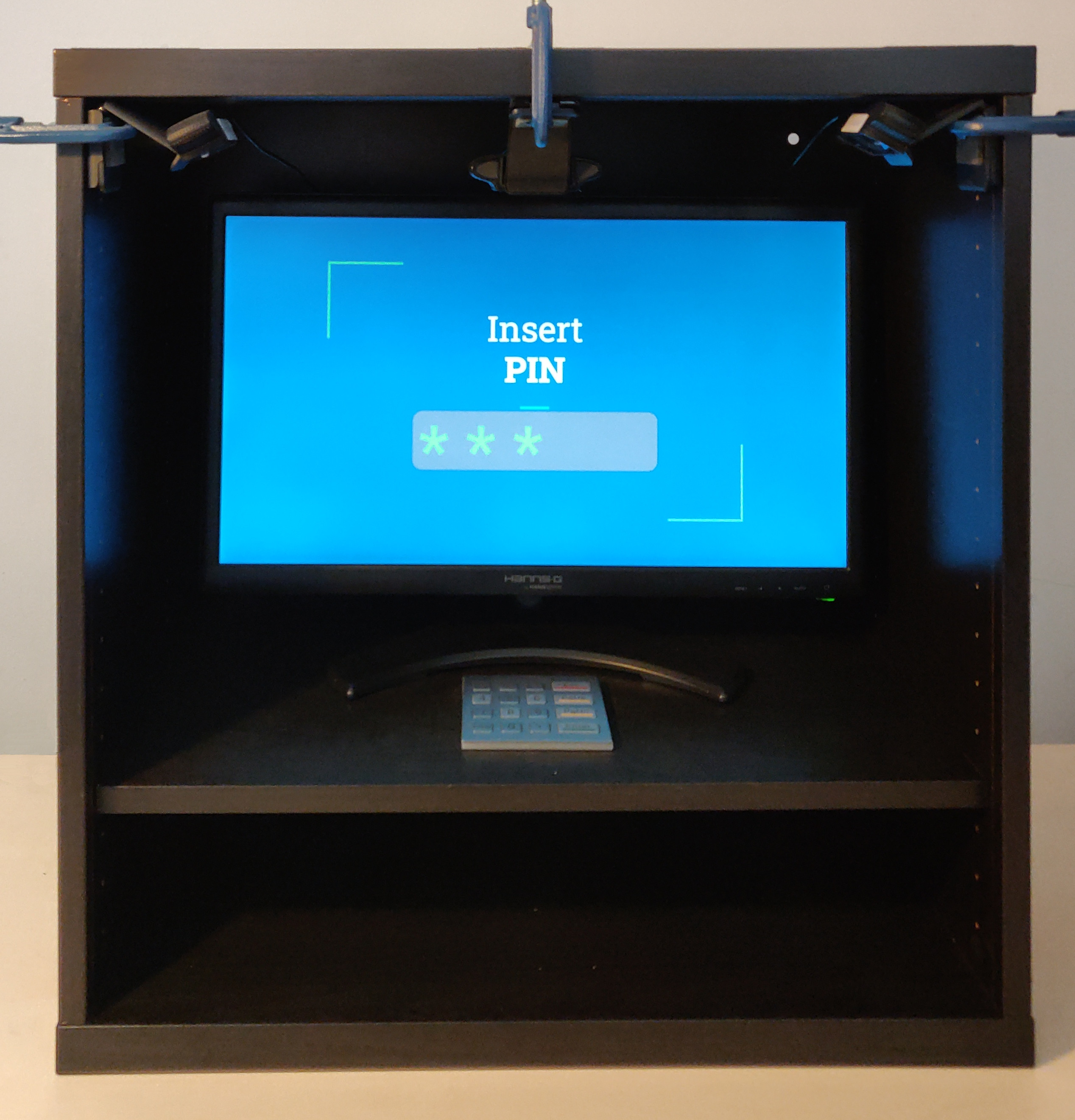}
	\caption{Our experimental setup. The cameras are visible but they can be hidden into the frame of an ATM. In all other aspects, we reproduced a common ATM layout in detail.}
	\label{fig:ATMTestbed}
\end{figure}

We performed two separated data collection campaigns on two different real-world ATM metal PIN pads: DAVO LIN Model \textit{D-8201 F}~\footnote{https://www.davochina.com/4x4-ip65-waterproof-industrial-metal-keypad-stainless-steel-keyboard-for-access-control-atm-terminal-vending-machine-p00103p1.html} (Figure~\ref{fig:PINPad1}) and Model \textit{D-8203 B}~\footnote{https://www.davochina.com/4x4-ip65-stainless-steel-numeric-metal-keypad-with-waterproof-silicone-cover-p00126p1.html} (Figure~\ref{fig:PINPad2}).
In particular, we report the following differences between the two PIN pads:
\begin{compactitem}
    \item Model \textit{D-8201 F} has a dimension of 100 mm $\texttt{x}$ 100 mm, while Model \textit{D-8203 B} has a metal surface of 92 mm x 88 mm and is contoured by rubber protection.
    \item The horizontal key spacing is 1 mm larger between each key in Model \textit{D-8203 B}.
    \item The keys of Model \textit{D-8203 B} are harder to press and slightly taller than Model \textit{D-8201 F}.
    \item For usability reasons, both the PIN pads emit a specific feedback sound (the same for all keys) when a key is pressed. The frequencies of the feedback sounds are 2\,900 Hz for Model \textit{D-8201 F} and 2\,500 Hz for Model \textit{D-8203 B}.
\end{compactitem}
For the data collection, we embedded the PIN pad into a simulated ATM (see Figure~\ref{fig:ATMTestbed}).
We chose the simulated ATM's size based on a real-world ATM~\cite{MONiMAX7600TA}. In particular, the simulated ATM has a width of 60 cm, a height of 64 cm, and a depth of 40 cm. At 15 cm of height from the frame's base, we inserted a shelf to position the PIN pad and the monitor. The height of the PIN pad from the ground is 110 cm. 
We used three \textit{Logitech HD C922 Pro} webcams anchored on the ATM's chassis to perform the video recording. A central webcam was placed 30 cm above the PIN pad, while the other two webcams were placed on the two top corners of the chassis 42 cm away from the PIN pad. The camera's maximum resolution is 1\,080p with an acquisition rate of 30 fps. We recorded the videos with a resolution of 720p and an acquisition rate of 30 fps.

\subsection{Data Collection}
\label{sec:DataCollection}



The first data collection involved 40 participants (age $38.23\pm11.43$, 24 male and 16 female). 
The second data collection involved 18 participants (age $29.50\pm5.74$, ten male and eight female). Both collections include right-hand participants only.
All the participants gave their approval to collect and use the data by signing informed consent. All the data have been anonymized and used by the authors of this paper for research purposes only.
Participants were asked to stand in front of the test ATM and cover the typing hand while entering the PIN during the experiment. The participants were left free to type as they pleased.
The goal is to emulate an ATM user that is hiding the PIN, preventing possible shoulder-surfing attacks.
Each participant typed 100 5-digits PINs randomly generated, divided into four sequences of 25 PINs. This split into four sequences has been performed to include short breaks in the experiments and prevent the participants from getting tired. The PINs were showed one at a time on the ATM screen: once a PIN has been entered on the PIN pad, the user had to press the enter button to move to the next PIN. We recorded a total of 5\,800 random 5-digit PINs, resulting in a balanced dataset per digit.
Since our study aims to reconstruct the PIN from the video sequence, regardless of the user's typing behavior and familiarity with the PIN or the PIN pad, we decided to randomize PINs rather than asking users to enter the same PIN multiple times. This approach generalizes the attack, which can be applied to mnemonic PINs and One-time Passwords (OTPs).
Moreover, we collected the environmental audio (exploiting the webcam microphone) and the keylogs of the PIN pad through the USB interface during the experiment. In particular, for each digit entered, we collect both the key down and key up events. We synchronized the video recordings with the timestamp of the key events. This information was collected to build the ground truth for the conducted experiments.
The dataset is available at~\url{https://spritz.math.unipd.it/projects/HandMeYourPIN}.

\subsection{Pre-processing Video}
\label{sec:Preprocessing}

Once the data acquisition phase is done, we need to pre-process the videos.
For each video frame, we applied the following steps: i) convert the video frames to grayscale; ii) normalize the input so that all pixel values lie in the range $[0, 1]$; iii) crop the frames by centering the PIN pad, cutting off the irrelevant part of the background;
(iv) resize the image to 250 $\texttt{x}$ 250 pixels.
After these steps, we applied a segmentation on each PIN video to obtain sub-sequences of frames corresponding to a single keypress (e.g., 5 sub-sequences for a 5-digit PIN). We extracted the keypress's timestamp from the recorded feedback sound of the PIN pad following the procedure explained in~\cite{cardaioli2020your}.
In particular, we filtered the audio signal using a band-pass filter, centered on the specific frequency of the feedback sound (i.e., 2\,900 Hz for Model \textit{D-8201 F} and 2\,500 Hz for Model \textit{D-8203 B}).
By identifying the peaks of the filtered signal, we could detect the timestamp of the target key (TK).
This allowed us to extract a set of frames in each TK neighborhood. 
For each TK, the maximum number of frames (full-neighborhood) consists of all the frames ranging from the key preceding the TK to the key following the TK. If the TK corresponds to the first digit of the PIN, we consider only the frames between the TK and the next keypress. Analogously, if the TK corresponds to the last digit of the PIN, the frames considered are only those between the TK and its previous keypress.
Since our model requires all input samples to have the same length, we decided to keep 11 frames for each sample. This value corresponds to the average number of frames in the full-neighborhood after removing the outliers over $3\sigma$. 
To keep the TK at the center of the frames' sequence, we decided to consider five frames preceding the target keypress and five frames succeeding it, for a total of 11 frames per sample (including the target frame).
There are three borderline cases: the TK is the first digit in the sequence, the TK is the last digit in the sequence, and the full-neighborhood has less than 11 frames. We apply black frame padding to keep the TK at the center of the sequence for these cases. In particular, if the TK is the first digit of the pin, five black frames are added at the head of the sequence, while if TK is the last digit of the PIN, we add five black frames at the end of the sequence. Finally, if there are not 11 frames in a sequence, we pad both the head and the tail (so that the TK is at the center).

\subsection{Machine Learning Setup}

For our experiments, we used a machine equipped with a CPU Intel(R) Xeon(R) E5-2670 2.60GHz, 128GB of RAM, and three Tesla K20m where each GPU has 5 Gb of RAM. 
To implement the machine learning models, we used Keras 2.3.0-tf (Tensorflow 2.2.0) and Python 3.8.6.

\subsection{Prediction Models}
\label{sec:PredictionModels}

Our approach aims to predict which key has been pressed on a PIN pad, exploiting only the video of a user covering the typing hand with the other hand.
Since we deal with sequences of images, we implemented a model using Convolutional Neural Networks (CNNs)~\cite{lecun1998gradient} and a Long Short-Term Memory (LSTM)~\cite{hochreiter1997long}. The CNNs perform spatial feature extraction for each frame of a sequence, while the LSTM exploits these features to extract temporal patterns for the whole sequence of frames. The output of the LSTM passes through a multilayer perceptron (MLP) and a final Softmax activation function layer with ten units (as there are ten digits).
This model is known in the literature as Long-term Recurrent Convolutional Network (LRCN)~\cite{donahue2015long}. In Keras~\cite{ketkar2017introduction}, such architecture can be implemented using the TimeDistributed wrapper throughout all the CNNs layers, which causes the same convolutional filters to be applied to all the timesteps (i.e., the frames) of the input sequence.

We split our dataset into train, validation, and test sets. Each set's size depends on the attack scenario and is discussed in detail in Section~\ref{sec:ExperimentalResults}.
We explored different hyperparameters by using the randomized grid search.
Based on a preliminary assessment, we set the ranges for specific hyperparameters (i.e., we limit the upper value for specific hyperparameters) to speed up the search.
In particular, for the CNNs, we tested [$3\texttt{x} 3, 6\texttt{x} 6, 9\texttt{x} 9$] kernel sizes. We also varied the number of convolutional layers in the range [1, $\ldots$, 4].
In the following dropout layer, we varied the dropout rates in the range [$0.01, 0.05, 0.1, 0.2$].
For the LSTM architecture, we varied the number of layers in the range [1, $\ldots$, 3], and the unit size in [$32, 64, 128, 256$]. 
We also assessed our network's performance using a Gated recurrent unit (GRU) instead of the LSTM. 
Finally, we examined the number of layers for the MLP in the range 1 to 4 and the number of units in the range $16, 32, 64, 128$. We tried two types of architectures for MLP: i) all the layers have the same number of units, ii) layers with decreasing number of units (funnel architecture), with every next layer having half the units of the previous one.

After a tuning phase, we selected a structure consisting of four convolutional layers (Conv2D in Keras) with ReLU activation functions, each followed by a pooling layer (MaxPooling2D in Keras). Three convolutional layers have a filter size of $3 \texttt{x} 3$, and one (the second one) has a filter size of $9 \texttt{x} 9$. Each pooling layer has a filter size of $2 \texttt{x} 2$. The number of filters in the convolutional layers doubles at each layer, starting from $32$ filters in the first layer, ending up at $256$ filters in the fourth layer. We added a dropout layer (dropout rate $0.1$) after the last pooling layer to prevent overfitting. The output is then flattened, preserving the temporal dimension to provide a sequence of temporal features to the following LSTM.
A single layer LSTM with $128$ units resulted in the best validation with a hyperbolic tangent activation function. Finally, for the MLP, we used four fully connected layers, with $64$ units each, followed by the Softmax activation layer with ten units (i.e., the number of classes we want to predict).
We used the categorical cross-entropy loss function and the stochastic gradient descent (SGD) optimizer. Finally, we set the model to evaluate the accuracy metric. We set the batch size to 16 and the learning rate to $0.1$.
We tested for 70 epochs since we found that the model always converged within this number of epochs. 
In Appendix~\ref{sec:architectures}, we provide additional details.
\new{Our experiments indicate that the classification task we conduct is relatively difficult, and one needs to use sophisticated deep learning architecture for good results. Still, we note that the architecture we use is in line with the state-of-the-art results for hand tracking problem~\cite{Lai18:cnnrnn, Isl19:hand_augmentation}. Finally, we observed significant changes in the performance depending on the specific hyperparameter choice, indicating a need for detailed tuning for the respective tasks.}

In a real-world context, it might not be possible to reproduce precisely the experimental conditions (e.g., the camera might be rotated/tilted slightly concerning the PIN pad, or the distance to the PIN pad might not be the same).
Thus, we also used data augmentation to generate synthetic measurements \new{(20\% of the training dataset)} that cover more scenarios to account for such issues.
In particular, we used the following video-based transformations:
\begin{compactitem}
    \item \textbf{rotation} for a maximum of $7\deg$ both clockwise and counterclockwise;
    \item \textbf{horizontal shift} for a maximum of 10\% of the width;
    \item \textbf{vertical shift} for a maximum of 10\% of the height;
    \item \textbf{zoom} between 0.9 and 1.1.
\end{compactitem}
\new{Synthetic samples were generated by randomly combining the transformation techniques listed above.}
We emphasize that data augmentation is also helpful as it makes the predictive model adaptable to different types of ATMs.

\section{Experimental Results}
\label{sec:ExperimentalResults}

In this section, we evaluate the performance of our approach for the three attack scenarios described in Section~\ref{sec:ThreatModel}. 
We adopted a user-independent split strategy since, in a realistic context, the attacker does not have labeled videos of victims entering PINs. In this way, we guarantee that videos from a participant appear only once among the three sets.
\new{Moreover, since we are interested in evaluating the PINs reconstruction accuracy, we removed all non-5-digit sequences entered by mistake by participants (i.e., the ''enter`` key was pressed after a sequence longer or shorter than 5-digits.) The removed non-5-digits sequences account for 2.2\% of the total PINs entered}.
We conducted the experiments on both 4-digits and 5-digits PINs.
To experiment on 4-digit PINs, we removed the last digit of each 5-digit sequence in our dataset.

We define that a PIN is covered when there is no direct view of the entered keys and their surrounding.
Still, we observed that some participants failed to obtain a satisfactory coverage level with the non-typing hand despite our instruction before starting the data collection.
Since this study aims to infer covered PINs, we decided to exclude the videos of participants that entered badly covered PINs from the validation and test sets. 
In this way, the validation and test sets consist of videos of covered entered PINs, while the training is composed of videos containing both covered and badly covered PINs. Note that badly covered PINs are still difficult to ``read'' by simply looking at the video, so we consider such data useful in building a training set. For the test set, we aim for the most difficult scenario where PINs are properly covered.
Under these assumptions, we ''blacklisted`` 16 participants that badly covered the PIN pad: 14 for the first data collection and two for the second data collection. These participants have been excluded from validation and test sets described in the below scenarios~\footnote{Results are in Appendix~\ref{app:additional_experiments}}.
In Figure~\ref{fig:CoveredAndUNCoveredPIN}, we provide an example of a badly covered PIN and a covered PIN.

To obtain a further indication of the quality of coverage and the difficulty of reconstructing a PIN by a human, we surveyed a random sub-sample of videos of covered PINs (Section~\ref{sec:Questionnaire}). 
Finally, there is a question of how to predict the PIN that is not guessed correctly from the first attempt. Since we consider each digit independently, we consider a mechanism where our best guess comprises of individual best guesses (for each digit). If that PIN is incorrect, we consider the digit where the two best guesses have the smallest difference. We change that digit to the second-best guess in our PIN, and we try again. The same procedure is repeated for the third attempt if the second PIN is wrong.

\begin{figure}
\centering
\subfloat[\emph{Badly covered PIN that we excluded from the validation and test tests.}\label{fig:UNCoveredPIN}]
{\includegraphics[width=0.45\linewidth]{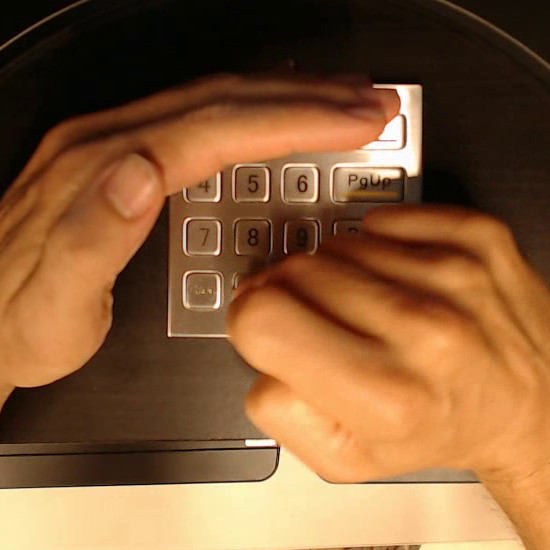}}\quad
\subfloat[\emph{Covered PIN, where there is no direct view of the pressed key and the surrounding digits.}\label{fig:CoveredPIN}]
{\includegraphics[width=0.45\linewidth]{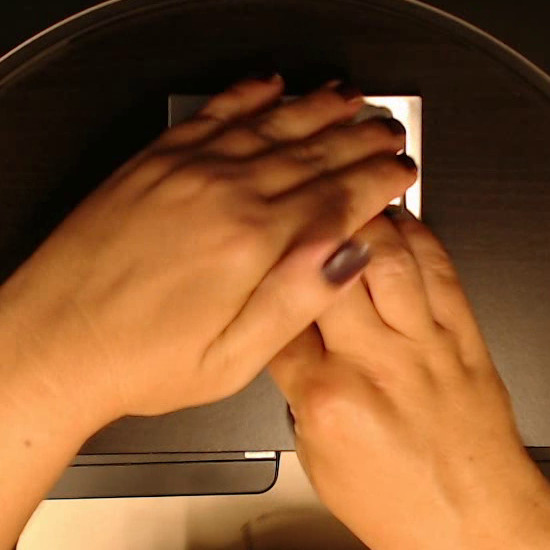}}
\caption{Badly covered vs. covered PINs.}
\label{fig:CoveredAndUNCoveredPIN}
\end{figure}



\begin{figure*}
\centering
\subfloat[\emph{True digit = 7\\ Pred = 7 (0.999), 4 (0.000), 8 (0.000)}\label{fig:digit_1}]
{\includegraphics[width=0.18\linewidth]{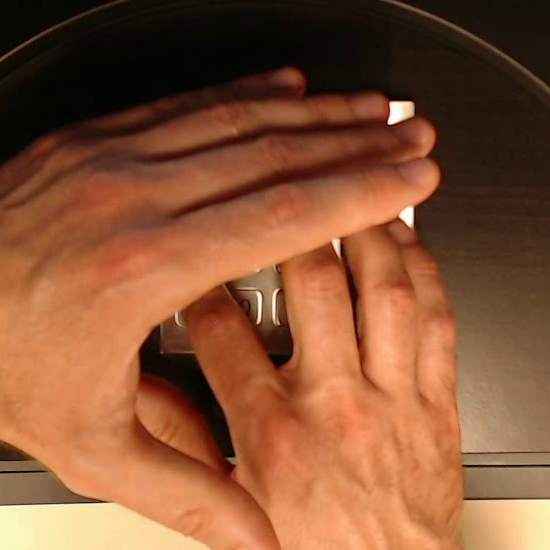}}\quad
\subfloat[\emph{True digit = 3\\ Pred = 3 (0.979), 2 (0.012), 6 (0.005)}\label{fig:digit_2}]
{\includegraphics[width=0.18\linewidth]{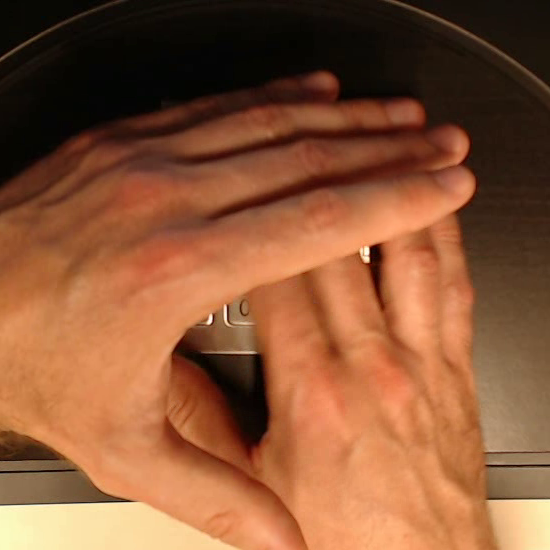}}\quad
\subfloat[\emph{True digit = 6\\ Pred = 6 (0.819), 9 (0.170), 8 (0.009)}\label{fig:digit_3}]
{\includegraphics[width=0.18\linewidth]{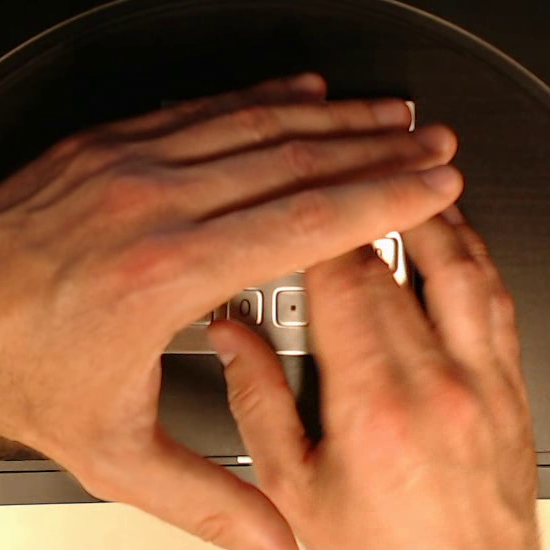}}\quad
\subfloat[\emph{True digit = 3\\ Pred = 3 (0.809), 2 (0.092), 5 (0.069)}\label{fig:digit_4}]
{\includegraphics[width=0.18\linewidth]{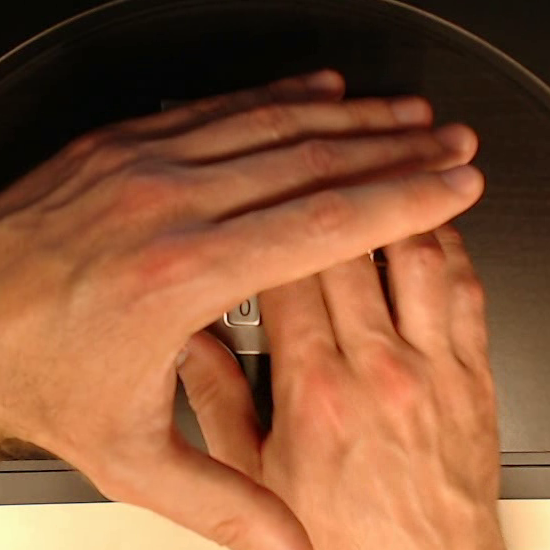}}\quad
\subfloat[\emph{True digit = 3\\ Pred = 2 (0.329), 3 (0.315), 6 (0.185)}\label{fig:digit_5}]
{\includegraphics[width=0.18\linewidth]{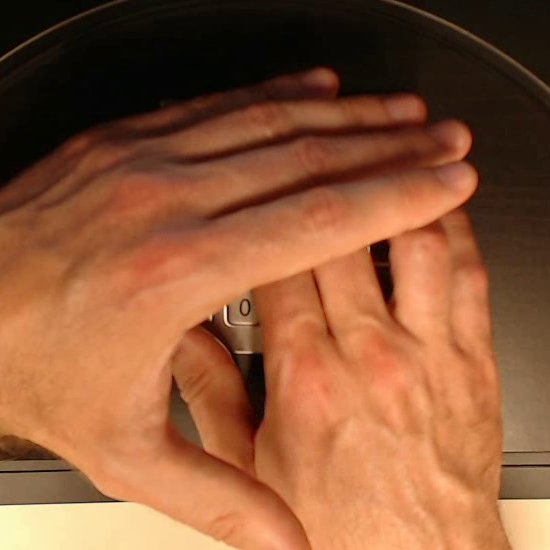}} 
\caption{PIN 73633 entered by a user in our test set in the \textit{Single PIN pad} scenario. Our algorithm suggests 73632 as the most probable PIN (probability = 21.32\%), 73633 as the second most probable PIN (probability = 20.43\%), and 73636 as the third most probable PIN (probability = 11.96\%).
The algorithm predicts the correct PIN in the second attempt.}
\label{fig:full_pin_example}
\end{figure*}

\begin{compactenum}
\item \textbf{Single PIN pad scenario}. To evaluate the scenario where the adversary knows the target PIN pad model and owns a copy, we considered only the first data collection composed of 40 participants.
We applied a user-independent split of the dataset in training, validation, and test sets with the proportions 80/10/10\%.
\item \textbf{PIN pad independent scenario.} In this scenario, the adversary trains the machine learning model on a PIN pad with a similar layout to the target one. 
This scenario occurs when the attacker cannot obtain the same PIN pad model to collect data.
Under these assumptions, we used for training and validation the first collected dataset (composed of 40 participants).
We included the videos from 35 participants in the training set and the remaining 5 participants' videos in the validation set.
We used the second collected dataset as the test set. We included only the videos of 16 out of 18 participants of the second data collection in the test set since two were in the group that badly covered the PIN pad.
\item \textbf{Mixed scenario.} 
This scenario corresponds to how the attacker owns both a copy of the target PIN pad and a PIN pad similar to the target one.
In this case, we merged the two collected datasets and applied a user-independent split in training, validation, and test sets with the proportions 80/10/10\%.
\end{compactenum}

We begin the discussion on results by providing an example of a successful PIN attack in Figure~\ref{fig:full_pin_example}. We consider the 5-digit PIN case and the \textit{Single PIN pad} scenario. We provide an image for each digit. We give the top three digits and the corresponding accuracy values. Notice how the first and second digits are predicted correctly with high probabilities. This happens as the person sets the hand to allow an easy start of typing. 
Already for the third digit, we observe a significant drop in the accuracy value for the best prediction. Still, the value is significantly larger than the second-best prediction, so there are no issues in getting the correct prediction.
This trend continues for the fourth digit and gets very pronounced for the last (fifth) digit. Indeed, the best guess is not correct anymore, but the second-best guess is correct (the difference in probability between those two guesses equals 0.014).

\begin{figure}[htp]
\centering
{\includegraphics[width=0.9\linewidth]{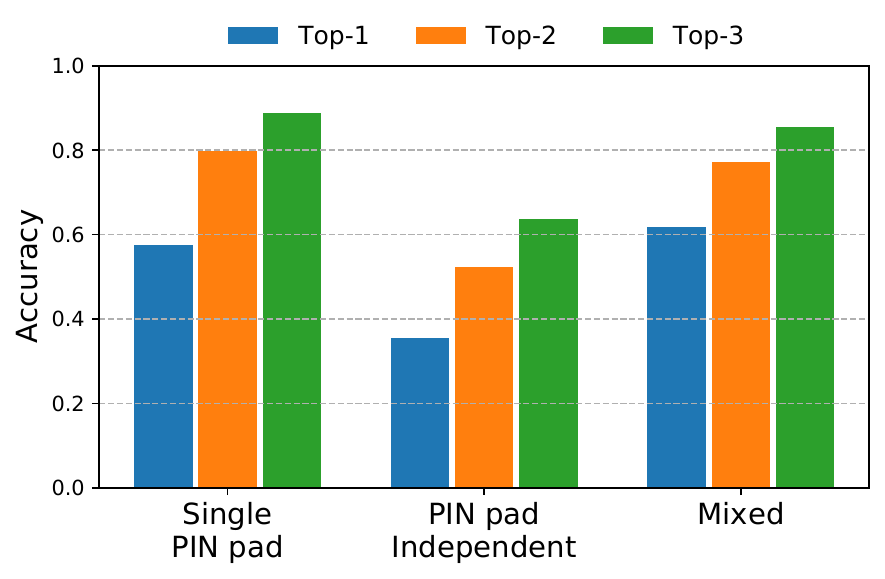}}\\
\caption{Single key accuracy of our algorithm for the three considered attack scenarios. Top-N means that we guessed the digit within the $N$ attempts.}
\label{fig:single_digit_accuracy}
\end{figure}

\begin{figure}
\centering
\subfloat[\emph{5-digit PINs.}\label{fig:5_digits}]
{\includegraphics[width=0.9\linewidth]{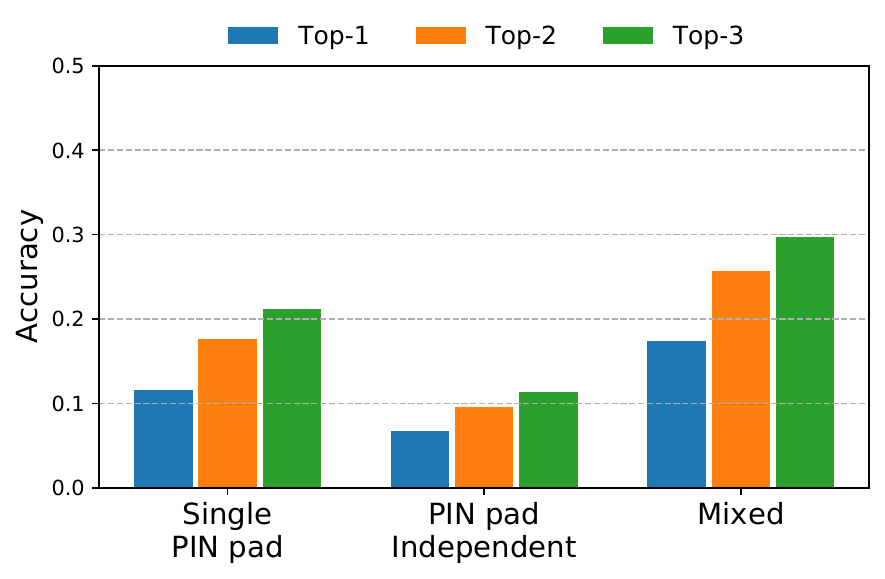}}\\
\subfloat[\emph{4-digit PINs.}\label{fig:4_digits}]
{\includegraphics[width=0.9\linewidth]{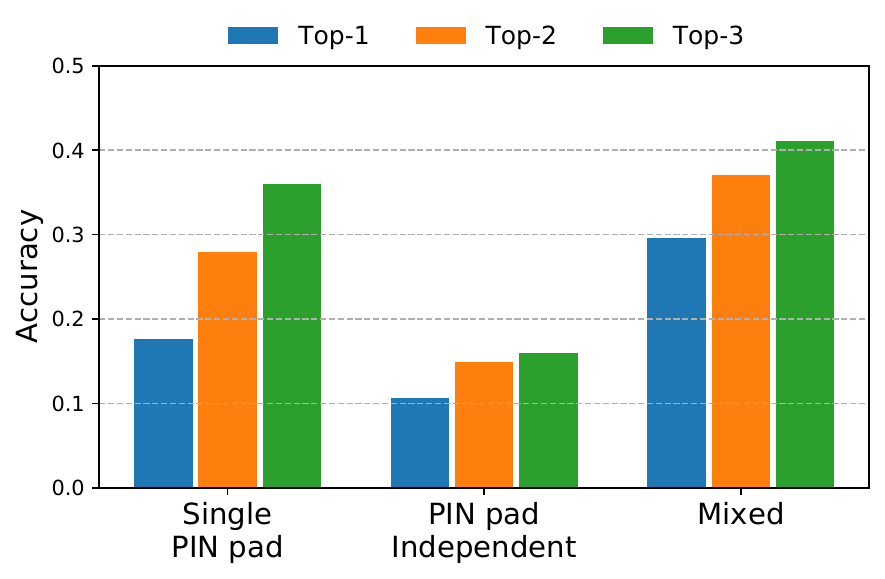}}\\
\caption{PIN accuracy of our algorithm in the three considered attack scenarios. Top-N means that we guessed the PIN within the $N$ attempts.}
\label{fig:4-5-digits-results}
\end{figure}

For all three scenarios, Figure~\ref{fig:single_digit_accuracy} shows the results for the single key accuracy, while Figure~\ref{fig:4-5-digits-results} reports the results considering 5-digit and 4-digit PINs.
Considering the single key accuracy (averaged over all digits), notice that even in the most difficult \textit{PIN pad independent} scenario, our Top-3 accuracy reaches 63.8\%, which is significantly higher than the result one would reach with random guessing (30\%). At the same time, the results for the \textit{Single PIN pad} scenario and the \textit{Mixed} scenario are rather similar, and the Top-3 accuracy reaches up to 88.7\%.
Interestingly, we observe somewhat better results for Top-2 and Top-3 accuracy for \textit{Single PIN pad} scenario than the \textit{Mixed} scenario, which is the opposite of the results for 4-digit and 5-digit settings. We hypothesize this happens as we consider independent digits as naturally, the best results happen when the training and test are done on the same device.
On the other hand, the \textit{Mixed} scenario gives slightly better results for the PIN reconstruction scenarios as we need to consider a sequence of PINs with the movement between digits. Then, having different devices in the training set allows (slightly) better generalization.

In Figure~\ref{fig:5_digits}, we observe that the most difficult case is when the attacker does not have access to the same keypad as used by the victim. There, the accuracy for the Top-3 case equals 11.4\%.
Having access to the same type of keypad improves accuracy in Top-3 to more than 20\%. Finally, considering the \textit{Mixed} scenario, we can improve the accuracy for Top-3 to almost 30\% (29.7\%). 
Next, in Figure~\ref{fig:4_digits}, we present results for 4-digit PINs. The results are significantly better than for the 5-digit scenario. The lowest accuracy happens for the Top-1 \textit{PIN pad independent} scenario setting and it equals 10.6\% (cf. 6.7\% for the 5-digit scenario). The highest accuracy reaches 41.1\% for the Top-3 accuracy in the \textit{Mixed} scenario. 

\begin{figure}[htp]
\centering
\subfloat[\emph{Layout of a generic PIN pad.}\label{fig:PINPadLayout}]
{\includegraphics[width=0.44\linewidth]{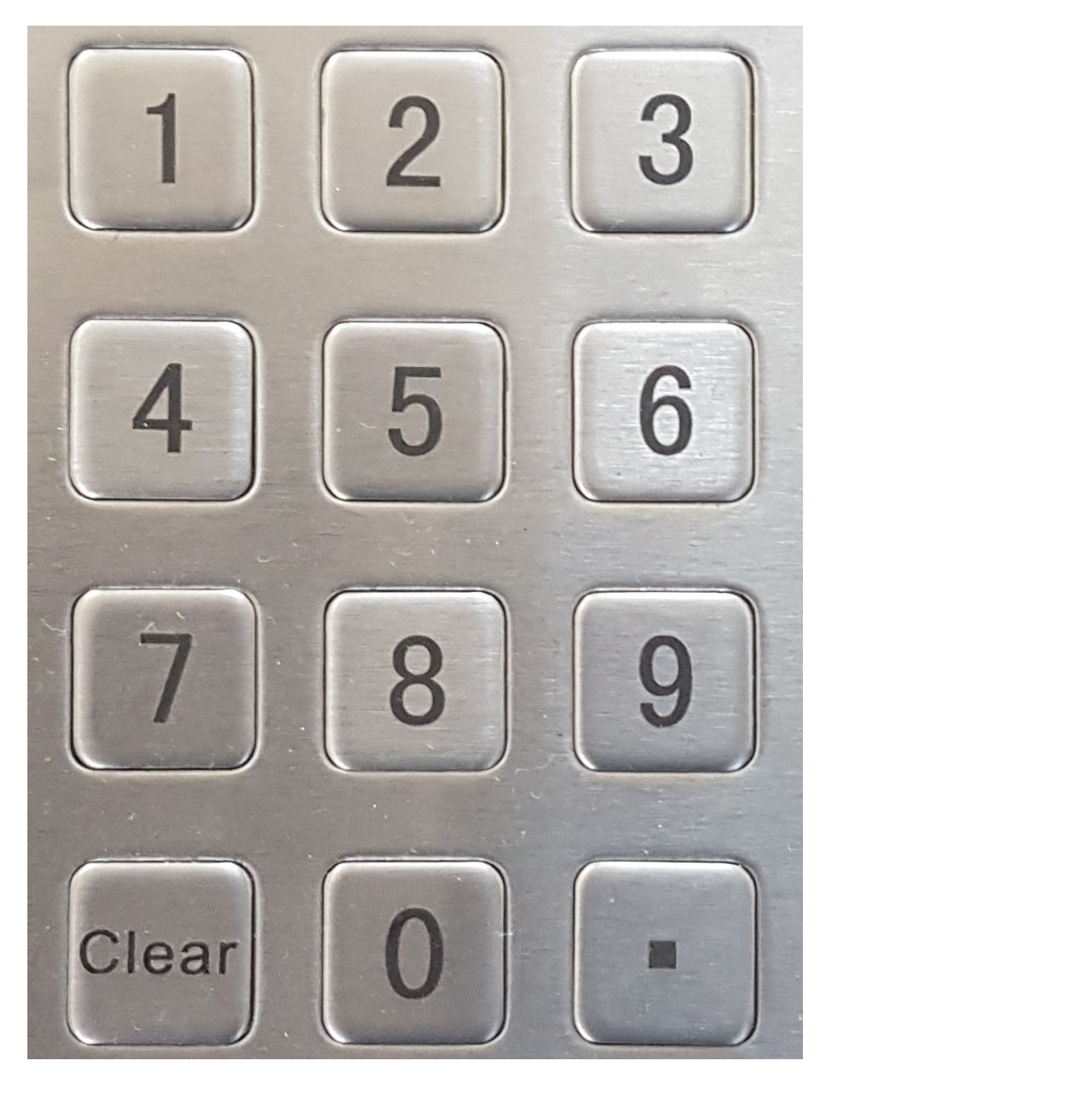}} \quad
\subfloat[\emph{Single PIN pad scenario.}\label{fig:D1onD1}]
{\includegraphics[width=0.44\linewidth]{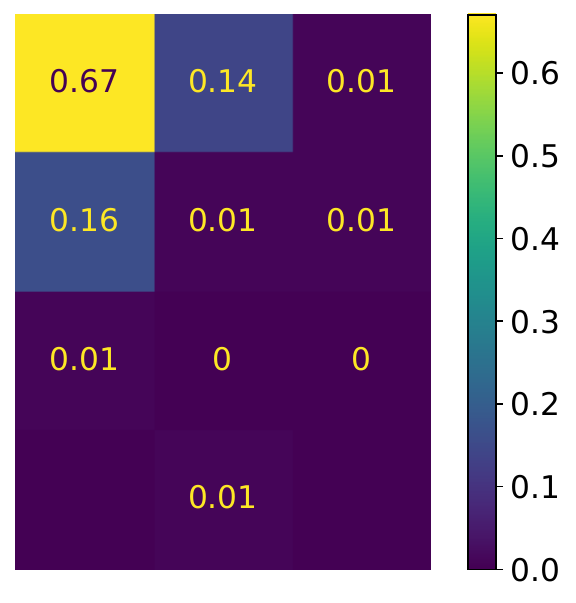}} \\
\subfloat[\emph{PIN pad independent scenario.}\label{fig:D1onD2}]
{\includegraphics[width=0.44\linewidth]{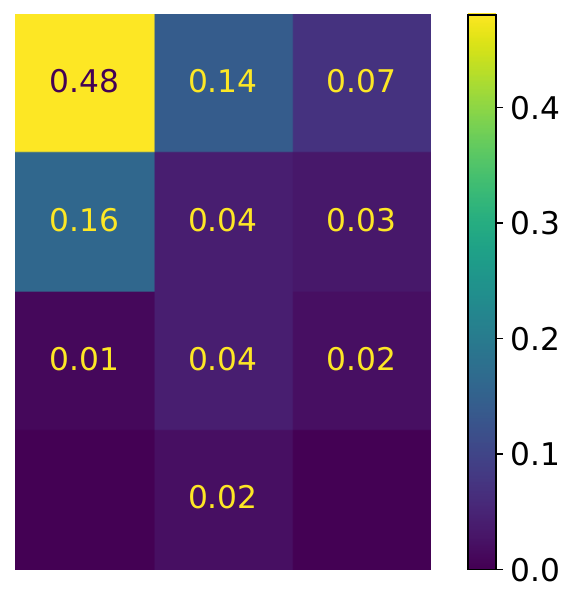}} \quad
\subfloat[\emph{Mixed scenario.}\label{fig:D1D2onD1D2}]
{\includegraphics[width=0.44\linewidth]{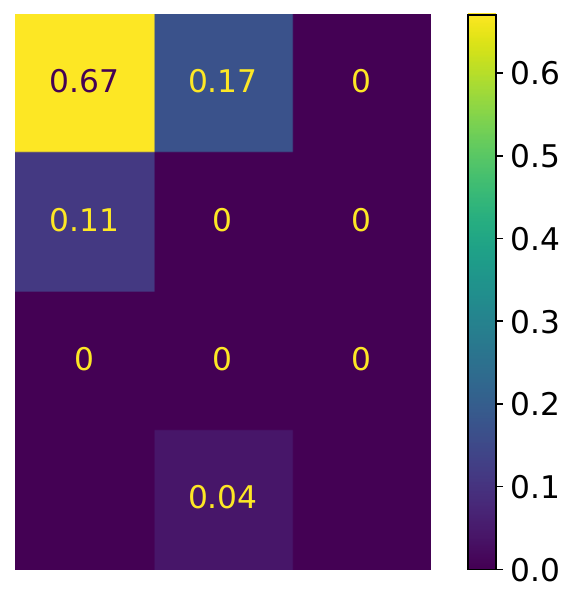}} \\
\caption{Digit $1$ predictions heat maps for the three considered attack scenarios.}
\label{fig:heat_map_1}
\end{figure}

In Figure~\ref{fig:heat_map_1}, we depict detailed results for the digit $1$. We selected this digit since heat maps for others look similar and exhibit similar dispersion.
First, in Figure~\ref{fig:PINPadLayout}, we show the PIN pad layout. Figure~\ref{fig:D1onD1} gives results for the \textit{Single PIN pad} scenario. Notice that the heat map indicates that guess 1 is the most likely one with 67\% probability. The digits 4 and 3 are recognized as the second and third best guess, respectively. Still, their probability is significantly lower. For the \textit{PIN pad independent} scenario, we observe that the probabilities are more spread over all digits, which comes at the expense of a lower prediction probability for the correct digit. The second and third best guesses maintain the probabilities, indicating that Top-3 guesses are sufficient to guess a large number of PINs in the most difficult scenario. Finally, Figure~\ref{fig:D1D2onD1D2} gives results for the \textit{Mixed} scenario, where we see that the best guess is on the level with the \textit{Single PIN pad} scenario. Interestingly, now the second and third best guesses are swapped compared to the previous scenarios. All the other digits have 0 or negligible probability of being the correct digit.
Appendix~\ref{sec:additional_results} provides additional results for the key accuracy.

\begin{figure*}[ht]
\centering
\subfloat[\emph{Side: hand resting on the side of the palm.}\label{fig:lateral_covering}]
{\includegraphics[width=0.23\linewidth]{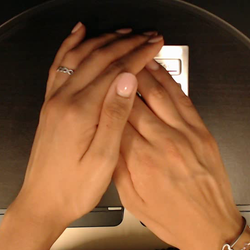}} \quad
\quad
\subfloat[\emph{Over: raised hand not touching the surface.}\label{fig:superior_covering}]
{\includegraphics[width=0.23\linewidth]{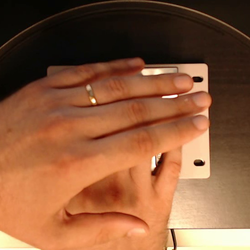}} \quad
\subfloat[\emph{Top: hand resting on fingers and vertically covering the PIN pad.}\label{fig:vertical_covering}]
{\includegraphics[width=0.23\linewidth]{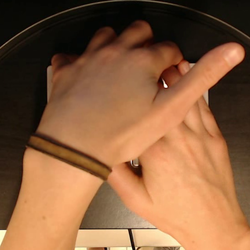}} 
\caption{Different covering strategies using the non-typing hand.}
\label{fig:covering_strategies}
\end{figure*}

Based on our results, we provide several observations that we believe generalize beyond these experiments:
\begin{compactitem}
\item Covering the PIN pad with the other hand is not sufficient to defend against deep learning-based attacks.
\item Portability aspect (keypad differences) is quite significant, and the attacker should obtain the same type of keypad for a high probability of success in attack.
\item There are three prevailing ways how users cover the typing hand: raised hand not touching the surface, hand resting on fingers and vertically covering the PIN pad, and hand resting on the side of the palm. The examples of all three covering strategies are shown in Figure~\ref{fig:covering_strategies}.
\end{compactitem}

\new{Finally, Table~\ref{tab:comparisonAttacks} provides a comparison between our attack and several unobtrusive attacks on 4-digit PINs from the literature~\cite{cardaioli2020your}. We divided the attacks according to the information that the attacker has: keystroke timing (KT), one digit of the victim's PIN (OD), and the thermal trace (TT) left on the PIN pad by the victim~\cite{abdelrahman2017stay}.
From the results, it is clear that our attack performs the best for all considered TOP-N accuracies.}

\begin{table}
\centering
\small
\begin{tabular}{@{}cccclll@{}}
\toprule
\multicolumn{4}{c}{\textbf{\begin{tabular}[c]{@{}c@{}}Attacker Information\\ Source\end{tabular}}} & \multicolumn{3}{c}{\textbf{\begin{tabular}[c]{@{}c@{}}4-digit PINs\\ TOP-N Accuracy (\%)\end{tabular}}} \\ \midrule
\textbf{\begin{tabular}[c]{@{}c@{}}KT\end{tabular}} & \textbf{\begin{tabular}[c]{@{}c@{}}OD\end{tabular}} & \textbf{\begin{tabular}[c]{@{}c@{}}TT\end{tabular}} & \textbf{\begin{tabular}[c]{@{}c@{}}Our Attack\end{tabular}} & \multicolumn{1}{c}{\textbf{TOP-1}} & \multicolumn{1}{c}{\textbf{TOP-2}} & \multicolumn{1}{c}{\textbf{TOP-3}} \\ \midrule
 &  &  &  & 0.01 & 0.02 & 0.03 \\
 & $\mathcal{x}$ &  &  & 0.10 & 0.20 & 0.30 \\
$\mathcal{x}$ &  &  &  & 0.02 & 0.35 & 0.72 \\
$\mathcal{x}$ & $\mathcal{x}$ &  &  & 3.02 & 3.72 & 4.36 \\
 &  & $\mathcal{x}$ &  & 3.76 & 7.52 & 11.28 \\
$\mathcal{x}$ &  & $\mathcal{x}$ &  & 15.54 & 27.79 & 33.63 \\
 &  &  & $\mathcal{x}$ & \textbf{29.61} & \textbf{37.06} & \textbf{41.12} \\ \bottomrule
\end{tabular}
\caption{Comparison of our attack with other unobtrusive attacks on ATM PIN pads. Note that we need to extract the frame for our attack, while for KT, one needs to use the timestamp, which is more precise information.}
\label{tab:comparisonAttacks}
\end{table}

\new{Appendix~\ref{app:additional_experiments} provides experiments where we: i) resize the images, ii) consider different camera positions, iii) consider setup without data augmentation, and iv) consider the training set that includes the blacklisted participants. 
Finally, we also provide experiments for the frame detection error (when the feedback sound is not properly synchronized).}


\section{Countermeasures}
\label{sec:countermeasures}

Different countermeasures could make the attack more difficult to succeed. For instance:
\begin{compactenum}
\item Longer PINs. This countermeasure would make the attack more difficult, as evident from the comparison for 4- and 5-digits PINs. This countermeasure would be relatively easy to support from a technical perspective. At the same time, it would have usability drawbacks as longer PINs take more time to type and are more difficult to remember.
\item Virtual and randomized keypad. Instead of using a mechanical keypad, one could consider using a touchscreen where the digits are randomized. More and more ATMs (but not PoS) have this feature, so implementing it would not be too difficult. Unfortunately, we believe this would seriously damage the usability aspect as people are accustomed to digits occurring in the natural sequence, and any changes would probably result in wrongly entered PINs.
\item Screen protectors. On many ATMs, there are already various types of screen protectors that occlude the typing hand. To maintain usability, many screen protectors are short and will not cover the whole typing hand. Making the screen protectors larger would impair usability as it will become more difficult for the user to read the keypad.  
This countermeasure is potentially not easy to deploy as it requires physical changes to the ATMs.
\end{compactenum}

\begin{table}
\centering
\small
\begin{tabular}{@{}lcc@{}}
\toprule
\textbf{Coverage} & \textbf{Key} & \textbf{PIN TOP-3} \\
\textbf{percentage} & \textbf{accuracy} & \textbf{accuracy} \\ \midrule
25\% & 0.54 & 0.22 \\
50\% & 0.55 & 0.22 \\
75\% & 0.50 & 0.17 \\
100\% & 0.33 & 0.01 \\ \bottomrule
\end{tabular}
\caption{PIN shield experiments.}
\label{tab:PINShielsPerformance}
\end{table}

\new{Next, we analyze how a PIN shield could affect the performance of our attack. We simulated the presence of the shield by applying a black patch to cover the PIN pad. In Table~\ref{tab:PINShielsPerformance} we report the performance of our attack in the \textit{Mixed} scenario, applying four different levels of coverage (Figure~\ref{fig:PINPadShield}, Appendix~\ref{app:additional_experiments}).
The coverage of the PIN pad is larger than the percentage shown in Figure~\ref{fig:PINPadShield} since the coverage given by the non-typing hand is not included in the given percentage.
The results show that our attack remains effective even when 75\% of the PIN pad is covered, while the performance decays significantly beyond this level of coverage. 
As such, it becomes clear that our deep learning attack uses information about the whole hand position and movement, and not only the tip of the fingers. Since the last row of the PIN pad has only one number (0), 100\% coverage has poor attack results not only because of hiding all the numbers on the keypad but due to hiding of proximal interphalangeal, metacarpophalangeal, and carpometacarpal joints of the fingers.
Thus, only PIN shields that offer full PIN pad coverage can be considered effective countermeasures to our attack.}

\new{We provide additional results with different covering strategies (\texttt{Side}, \texttt{Over}, and \texttt{Top}) in Appendix~\ref{app:additional_experiments}. Those results again show that covering the PIN pad from the \texttt{Side} gives insufficient protection. On the other hand, using the \texttt{Over} strategy significantly decreases the key accuracy and PIN accuracy.}

\section{Deep Learning vs. Humans}
\label{sec:Questionnaire}
If an attacker has direct visibility of the PIN pad, reconstructing a PIN from a video can be considered a trivial task. One of the classic countermeasures to the so-called shoulder-surfing attacks is to cover the hand entering the PIN with the non-typing hand. In this way, the victim obstructs the attacker by removing the direct visibility of the keypad. We designed a questionnaire to evaluate how much the covering with the non-typing hand effectively prevents the PIN reconstruction.

\subsection{Methodology}
The questionnaire consists of 30 videos of people entering 5-digit PINs by covering the PIN pad with the non-typing hand as we noticed that for longer questionnaires, the participants' attention significantly goes down toward the end.
For each video, the participants had to indicate the three most likely PINs in their opinion.

To assess human and model performance on both the PIN pads, we decided to use the test set of the \textit{Mixed} scenario (i.e., the only one including both PIN pads).
Since the test set was balanced in terms of samples per user, we randomly selected five PINs for each of the six users in the test set.
We extracted 30 videos corresponding to the selected PINs from our dataset. We kept the original resolution of 720p and the original audio track containing the feedback sound emitted by the PIN pad for each video. The feedback sound helps the participants to recognize when a digit is entered.
To avoid bias in the answers, we randomized the order of the videos in the questionnaire. Moreover, the participants were free to modify all their answers until the final submission.
We did not apply any particular restriction to the participants during the filling of the questionnaire.
In particular, there were no time restrictions to complete the task. The participants could freely apply the strategy they prefer to infer the PIN (e.g.,  write down the digits, pausing the video, restart the video any number of times, use the slow-motion option). Finally, we provided the users with the layout of the PIN pad.

\new{To evaluate if people with specific knowledge about the task achieve a better performance, we pre-trained a group of participants. Specifically, we provided participants with a new set of 20 videos of users typing PINs by covering the PIN pad with the non-typing hand and the corresponding typed PIN. To make the training more effective, we decided to provide participants with videos of users included in the questionnaire (none of the videos are present in both training and questionnaire). Additionally, the questionnaire had suggestions on what to pay special attention.
For a participant to be considered trained, the complete viewing of all 20 videos is required.
In addition, trained participants could also watch the training videos while filling the questionnaire.}

\subsection{Evaluation and Discussion}

\new{A total of 78 distinct participants took part in our questionnaire experiment. In particular, 45 participants (14 female age $34.1\pm10.4$ years and 31 male age $29.7\pm8.3$ years) completed the experiment without any training, while 33 participants (10 female age $29.1\pm3.3$ and 23 male age $29.3\pm5.6$) completed the experiment after the training session.
None of the questionnaire participants took part in the two data collections described in Section~\ref{sec:DataCollection}.}

The proposed questionnaire's goal is twofold: i) investigate how effective the hand coverage is in preventing a PIN from being inferred by a human, and ii) compare the performance of our deep learning approach with that of a human.
Although the coverage of the PIN pad provides an obstacle to the immediate identification of the typed PIN, a human can exploit various information (both local and global) to reduce the probability space about where to look for the entered PIN:
\begin{compactitem}
\item Knowing the keys' spatial positioning thanks to the given layout of the target PIN pad.
\item Understanding which finger pressed the key from the movements of the hand.
\item Evaluating the topological distance between two consecutive keys from the feedback sound emitted by the PIN pad. Specifically, two topologically close keys have temporally close sound feedback~\cite{cardaioli2020your}.
\item Excluding keys based on the non-typing hand coverage.
\item Guessing the finger position based on the hand displacement between the insertion of a key and the next one.
\item Deducing the fingers' position of the covered hand.
\end{compactitem}

\begin{figure}[ht]
\centering
\subfloat[\emph{Humans.}\label{fig:UserCM_4}]
{\includegraphics[width=0.47\linewidth]{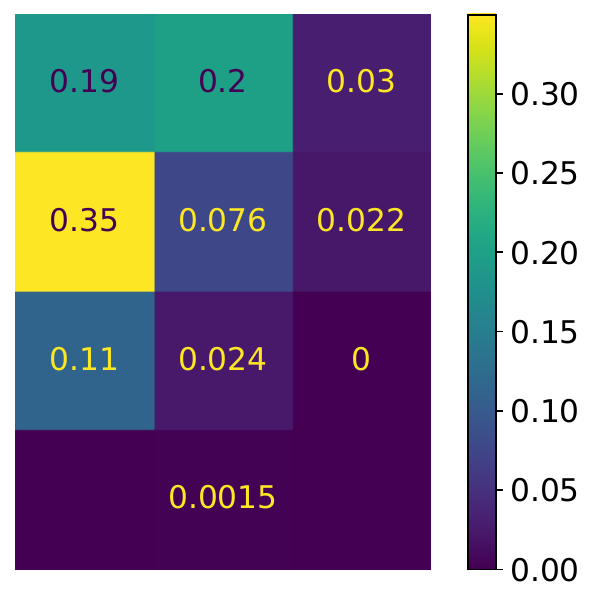}} \quad
\subfloat[\emph{Mixed} scenario model.\label{fig:RecalculedCM_4}]
{\includegraphics[width=0.47\linewidth]{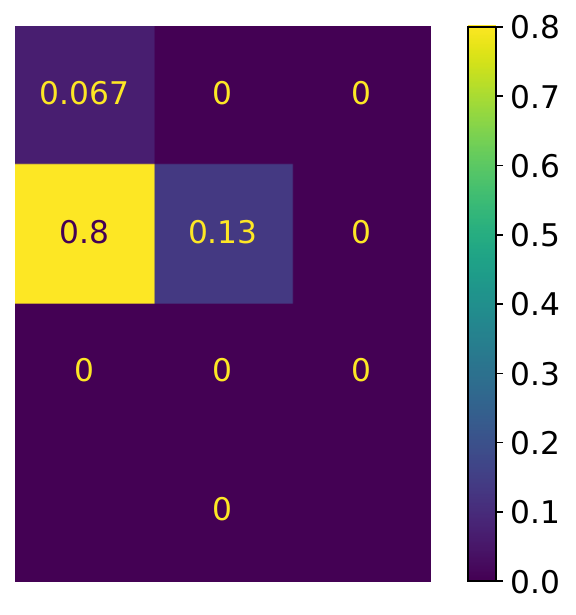}} \quad
\caption{Digit $4$ predictions heat maps for the videos included in the questionnaire. \new{We report an example from non-trained humans, since the heat maps for both non-trained and trained human are similar.}}
\label{fig:survey_heat_map_single_digit}
\end{figure}

Although a human can exploit this information, the PIN pad coverage still partially prevents PIN reconstruction. In particular, the participants in our questionnaire could reconstruct on average (of both trained and non-trained humans) only 4.49\% of the PINs entered in the videos on the first attempt and 7.92\% within three attempts. The performance increasing between Top-1 and Top-3 accuracy suggests a certain ability in estimating the neighborhood of the keys pressed. This ability is also highlighted in Figure~\ref{fig:UserCM_4}, where the probability distribution shows how the error decreases with the increase of the topological distance from the target key. The heat maps for other keys look similar and exhibit similar dispersion.

\begin{figure}[ht]
\centering
\includegraphics[width=0.9\linewidth]{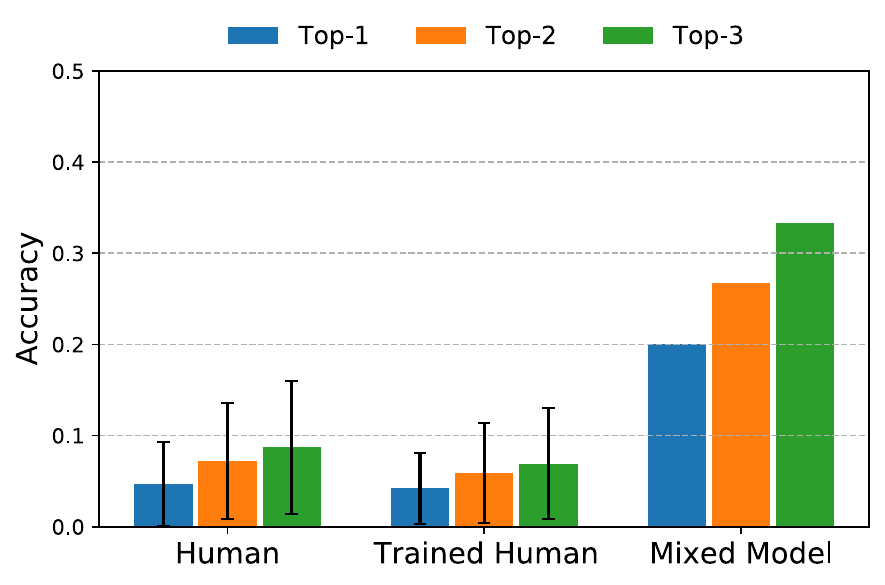} 
\caption{Comparison between human (non-trained and trained) and deep learning model performance in the sub-set of videos included in the questionnaire. Top-N means that participants guessed the PIN within the $N$ attempts.}
\label{fig:human_vs_model}
\end{figure}


Unlike humans, our algorithm focuses on target key classification and then reconstructs the entire PIN sequence.
To compare the model's performance to that of humans on the same task, we evaluated our algorithm's accuracy on the videos included in the questionnaire. Recall that the questionnaire's videos are a sub-sample of the \textit{Mixed} scenario test set, and therefore were not used in the model training phase.
As reported in Figure~\ref{fig:human_vs_model}, our model performs better than humans in all Top-N accuracy scenarios. To evaluate if our algorithm performance and humans' performance in reconstructing 5-digit PINs are statistically different, we applied a series of Chi-square tests~\cite{mchugh2013chi}.
\new{The Chi-square test resulted significant for all Top-1 ($\chi^2 = 14.19, p<0.001$), Top-2 ($\chi^2 = 15.84, p<0.001$), and Top-3 ($\chi^2 = 21.37, p<0.001$) accuracy values for non-trained humans. In particular, our model outperforms humans showing a four-fold improvement in reconstructing a PIN in three attempts. Similarly, for trained humans, the Chi-square test resulted significant for all Top-1 ($\chi^2 = 16.12, p<0.001$), Top-2 ($\chi^2 = 20.83, p<0.001$), and Top-3 ($\chi^2 = 28.88, p<0.001$) accuracy values.}





\new{This result comes from the difference in performance in the classification of single keys. The human average accuracy (considering both human data collections) on single key classification equals 0.351, approximately half compared to the model key accuracy of 0.687.
The comparison of Figures~\ref{fig:RecalculedCM_4} and~\ref{fig:UserCM_4} shows how the error in identifying a digit is significantly higher for humans, justifying why the increase in Top-2 and Top-3 PIN accuracy is greater for our algorithm.
Finally, comparing trained and non-trained humans, the Chi-square test reported no significant differences with $p>0.1$ for all Top-1, Top-2, and Top-3 accuracy values. This means that training does not improve a human's ability to identify a PIN within three attempts. Potentially, either a longer training could be required, or additional feedback from an expert should be provided to improve the performance.}
Appendix~\ref{sec:additional_results} provides additional results for the comparison between our deep learning model and human performance.

\section{Related work}
\label{sec:related_workd}

Side-channel attacks specifically target the information gained by the implementation of a system~\cite{dpa_book}. Most of the time, these attacks exploit channels like sound~\cite{10.1007/s00145-015-9224-2}, timing~\cite{kocher1996timing}, power consumption~\cite{kim2019make}, and electromagnetic emanations~\cite{DBLP:conf/ndss/BhasinCHJPS20} to learn the system's secrets in use.
In~\cite{kocher1996timing}, the authors managed to crack RSA keys by carefully timing the operations performed by the key-generating algorithm. Another example of a timing attack is reported in~\cite{song2001timing}, where the authors measured the timing between keystrokes in interactive SSH sessions in an attempt to retrieve the typed passwords.

Human behavior can also be defined as a side-channel of a system, especially if the analyzed behavior directly results from the system's requirements.
In~\cite{balzarotti2008clearshot}, the authors analyzed the hand movements of people typing on a keyboard and, by using basic computer vision techniques, they tried to reconstruct the text being typed. In~\cite{shukla2014beware}, the authors again analyzed the finger motion during the PIN-entry process on smartphones. They showed that 50\% of the 4-digit PINs could be retrieved in just one attempt.
Different from our work, where the target of the attack is a physical PIN pad, in~\cite{shukla2014beware}, the attackers could also exploit more information. In particular, the users typed the PIN using only one finger, and the attacker knew the finger the users are typing. The different contexts and assumptions make the works substantially different.
In~\cite{sun2016visible}, the authors presented a side-channel attack on tablets, consisting of analyzing the backside movements of the tablet itself to infer what is being typed by the victim. To do so, they selected some peculiar features of the backside of the tablets (e.g., logos, side-buttons) and analyzed their movement throughout the frames to understand what area of the virtual keyboard is being pressed.
\new{Similarly, in~\cite{ye2017cracking}, the authors presented an attack to infer the pattern lock of mobile devices from videos. Different from our approach, in~\cite{ye2017cracking}, the attacker required a vision of the user’s fingertip while drawing the pattern and a part of the device.}

PIN and PIN pad attacks represent a branch of side-channel attacks that exploit information leakage from keyboards and numeric keyboards (i.e., PIN pads) to infer what the victim has typed (e.g., passwords or PINs).
In this context, some works focused on exploiting the heat transferred from the hand to the keypad when the victim enters the PIN or password~\cite{mowery2011heat, kaczmarek2019thermanator}. The attacker points a thermal camera to the keypad as soon as the victim has finished entering the PIN. The thermal image shows which keys have been pressed and even highlights the order in which the victim pressed them.
The main advantage of this attack is that it does not require the attacker to do anything while the victim is typing the PIN. On the other hand, the attacker must act quickly (i.e., within seconds) for a higher success rate as the heat on the keypad rapidly fades away. Another drawback of the attack is that its effectiveness depends on the keypad's material (e.g., metal PIN pads completely nullify the attack because of their high thermal conductivity).

Timing attacks against PINs represent another type of side-channel attack against this authentication method. In the scenario presented in~\cite{balagani2019pilot}, the attacker recorded the screen of an ATM while the victim is entering the PIN. When analyzing the recorded video, the attacker exploited the PIN masking symbols appearing on the ATM screen to extract timing information about the keystrokes. The attacker used predictive models to infer which keys were most likely typed by the victim, starting from the deduced inter-keystroke timing.
In~\cite{cardaioli2020your}, the authors used the ATM's sound whenever a button is pressed. ATM's sound must be independent of which button is being pressed (i.e., a generic feedback sound). This consideration means that one feedback sound will not help the attacker. However, the sound gives enough information to extract a timestamp of the keys being pressed. Moreover, in~\cite{cardaioli2020your, liu2019human}, the authors showed how combining timing, acoustic, and thermal information can significantly reduce the number of attempts to guess a PIN (e.g., 34\% of 4-digits PINs are recovered in three attempts).
These attacks need to be reevaluated from a feasibility perspective in a real-world setting.
In particular, as shown in~\cite{kaczmarek2019thermanator}, the heat signature is dissipated abruptly by metal PIN pads. The lack of this information limits the performance of the attacks presented in~\cite{cardaioli2020your, liu2019human}, reducing the probability of guessing a 4-digit PIN in 3 attempts to 5\%. 

\textit{Our work shows several advantages over the state-of-the-art in ATM PIN inference.
To the best of our knowledge, we are the first to investigate the security of hand covering protection methods for ATM's PIN entering. Further, our method shows a significant improvement in reconstructing the PIN compared to previous work on metal PIN pads, reaching 41\% of success in reconstructing 4-digit PINs in three attempts (and correctly guessing every third PIN in the first guess).}



\section{Conclusion}
\label{sec:conclusions}
This paper proposed a deep learning attack on PIN mechanisms reaching high accuracy even when the user covers the PIN to be entered. Our attack leverages the information from the hand position but also hand movements while entering the PIN. 
Our attack works in the profiling setup where the attacker uses a copy of the keypad to train the deep learning model and then attacks a different device while the victim is entering the PIN.
For a 4-digit PIN, our attack reaches an accuracy of more than 40\%, making it practically applicable and more powerful than the attacks from the related works.

Our data collection phase involved 58 persons, and our questionnaire involved 78 participants. While this required a significant effort and several months of data acquisition, one could still consider the datasets too small to allow general conclusions.
Next, our analysis considered only two types of keypads. While most keypads do not have significant differences, including more keypad models in our analysis would be interesting.
Additionally, there are several potential sources of bias in our data collection phase. While we managed to get a relatively good male and female participants ratio, we notice that data is skewed from several perspectives. Unfortunately, this was not possible to avoid as the participation was voluntary~\footnote{The 2021 COVID-19 situation made data acquisition more challenging as participants needed to be in our lab during the data acquisition.}.
\begin{compactenum}
\item Our dataset has users ranging from 24 to 50 years. While this provides good variety, it would be good if it included older people. Still, we do not expect any difficulties in running our attack. We consider it even somewhat easier as we noticed older people make more significant hand position adjustments when entering the PIN.
\item Our analysis includes only right-handed persons. We do not expect any issues due to the dataset's limitations as we use a camera positioned in the center. Still, we expect the attack to be more difficult when attacking left-handed persons if the training set does not contain such examples. Finally, from the real-world practicality, there are approximately 90\% of right-handed persons vs. 10\% left-handed persons~\cite{f9605b8a435a4851a329948707b4e88d}, so our attack generalizes for the dominant part of the population.
\item All participants were Caucasians. We expect our attack will have difficulties working for people from other races. Still, this can be alleviated by expanding the training set to include more racial diversity.
\end{compactenum}


Possible future work includes:
\begin{compactenum}
\item In our data collection phase, we allowed the users to select their covering strategies. Based on the current results, it would be interesting to explore if modifications in how the user covers the PIN would allow more protection.
\item We noted several potential sources of bias in our data collection phase. Including participants from other races and left-handed persons would allow us to make more general conclusions.
\item To avoid the need that the attacker should have different keypads, it would be beneficial to assess whether some more straightforward solution like a paper copy of the keypad would suffice (at the expense of losing information about the keypress sensitivity).
\item It would be interesting to investigate if it is possible to extract the timestamp directly from the video (when a person clicks a button, there is a specific movement).
\end{compactenum}



\bibliographystyle{plain}
\bibliography{bibliography}

\appendix
\section{Neural Networks Additional Info}
\label{sec:architectures}



\begin{figure*}[htp]
\centering
\subfloat[\emph{Single PIN pad scenario. We included 4 participants in validation, corresponding to 400 digits.}\label{fig:D1_D1_accuracies}]
{\includegraphics[width=0.32\linewidth]{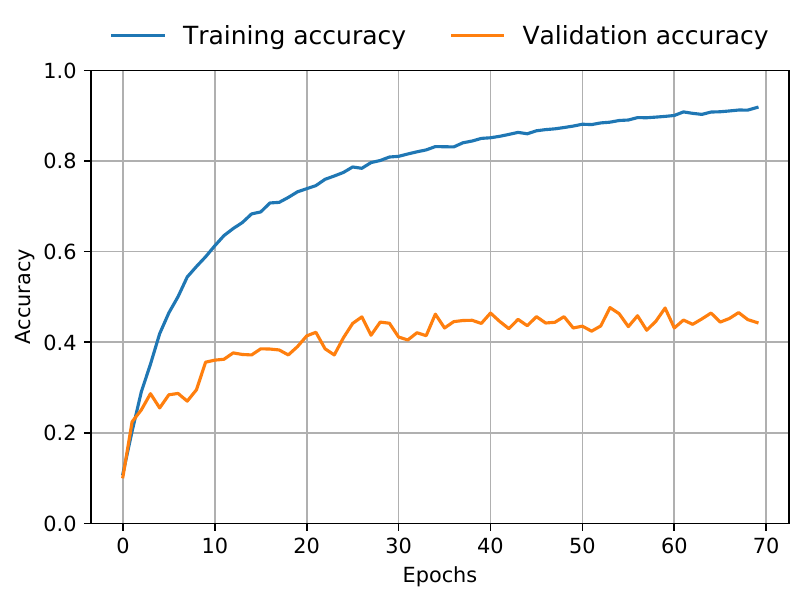}}
\subfloat[\emph{PIN pad independent scenario, We included 5 participants in validation, corresponding to 500 digits.}\label{fig:D1_D2_accuracies}]
{\includegraphics[width=0.32\linewidth]{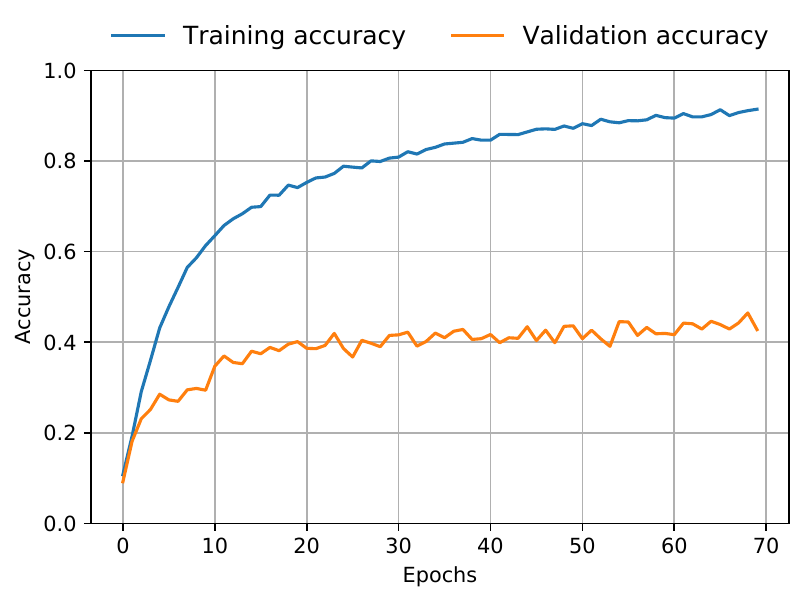}}
\subfloat[\emph{Mixed scenario. We included 6 participants in validation, corresponding to 600 digits.}\label{fig:D1D2_D1D2_accuracies}]
{\includegraphics[width=0.32\linewidth]{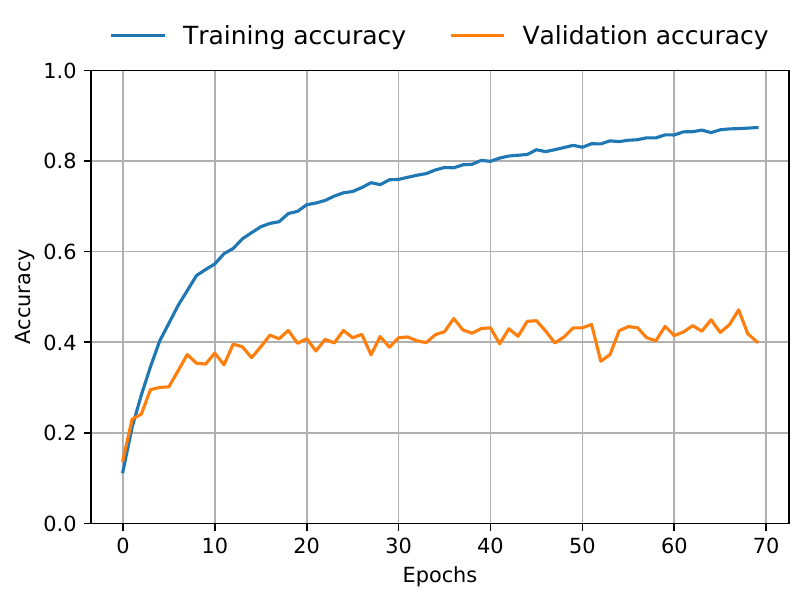}}
\caption{Training and validation accuracy for our three scenarios.}
\label{fig:validation_results}
\end{figure*}

In Figure~\ref{fig:validation_results}, we show the training and validation accuracy for the three models selected after the random grid search. In the \textit{Mixed} scenario, the validation accuracy grows faster than in \textit{PIN pad independent} scenario and \textit{Single PIN pad} scenario, reaching faster the plateau. Indeed, in the \textit{Mixed} scenario, the validation accuracy stabilizes after 20 epochs, while we require more than 35 epochs for the other scenarios. This difference can be linked to a larger training size and a higher variance in the samples since the \textit{Mixed} scenario is the only one to include videos from both PIN pads in the training phase.

\new{Next, we report some statistics about the training execution times for the three scenarios we consider.
\begin{compactitem}
\item \textbf{Single PIN pad} scenario: the training set is composed of 32 participants, corresponding to 16\,000 samples of 11 frames each. Our model takes 1\,577 seconds to complete an epoch (i.e., approximately 34 hours to complete the entire training phase).
\item \textbf{PIN pad independent} scenario: the training set is composed of 35 participants, corresponding to 17\,500 samples of 11 frames each. Our model takes 1\,598 seconds to complete an epoch (i.e., approximately 34 hours to complete the entire training phase).
\item \textbf{Mixed} scenario: the training set is composed of 46 participants, corresponding to 23\,000 samples of 11 frames each. Our model takes 2\,240 seconds to complete an epoch (i.e., approximately 46 hours to complete the entire training phase).
\end{compactitem}}

\section{Key Accuracy Analysis}
\label{sec:additional_results}

\begin{figure*}[htp]
\centering
\subfloat[\emph{Single PIN pad} scenario.]
{\includegraphics[width=0.3\textwidth]{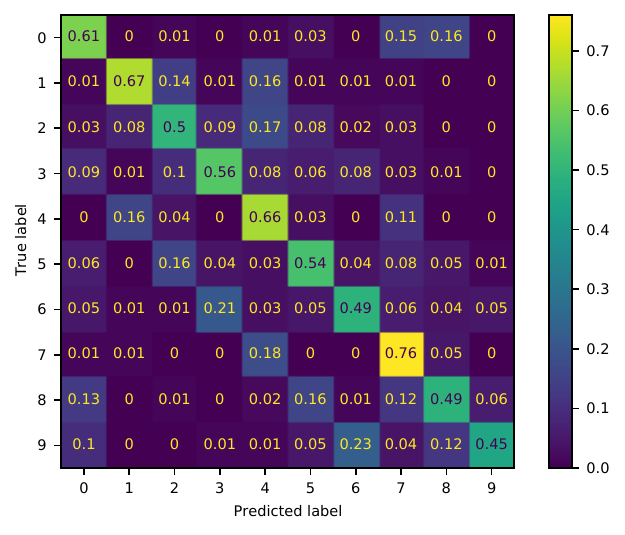}}\quad
\subfloat[\emph{PIN pad independent} scenario.]
{\includegraphics[width=0.3\textwidth]{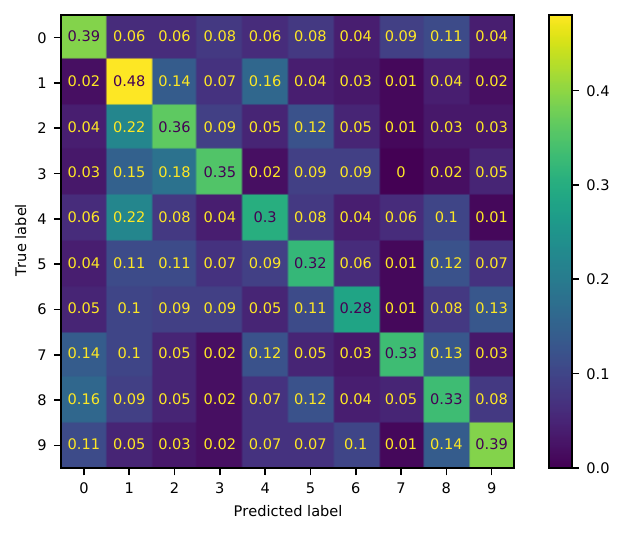}}\quad
\subfloat[\emph{Mixed} scenario.]
{\includegraphics[width=0.3\textwidth]{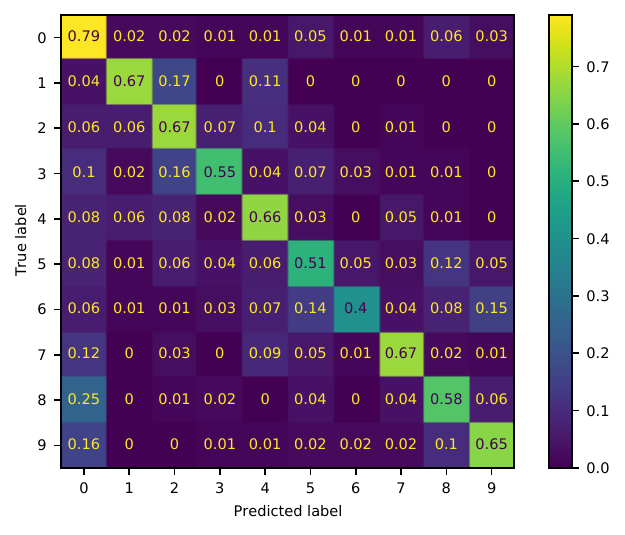}}\quad
\caption{Confusion matrices of key predictions (predicted labels) vs. true values (true labels) for our three scenarios.}
\label{fig:ConfusionMatrixies}
\end{figure*}

\new{In this section, we provide further analysis on the key accuracy for our attack. Figure~\ref{fig:ConfusionMatrixies} highlights that the accuracy on a single key is worse in the \textit{PIN pad independent} scenario. Although the performance is considerably lower than the other two scenarios in Top-1 accuracy, it is interesting that the error dispersion affects the keys topologically close to the target one.}

\begin{figure*}[htp]
\centering
\subfloat[\label{fig:RecalculedCM}][\emph{Recalculated confusion matrix for our algorithm (\textit{Mixed} scenario).}]
{\includegraphics[width=0.3\textwidth]{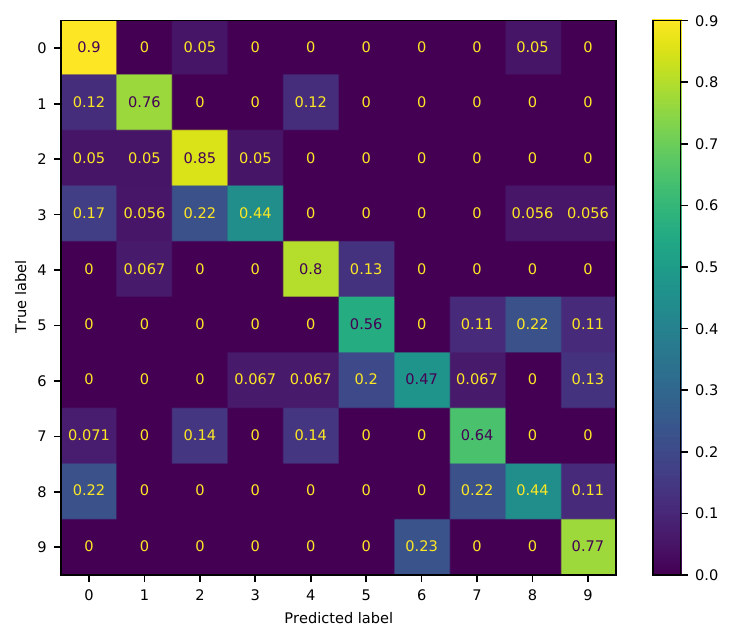}} \quad
\subfloat[\label{fig:UserCM}][\emph{Confusion matrix for humans.}]
{\includegraphics[width=0.3\textwidth]{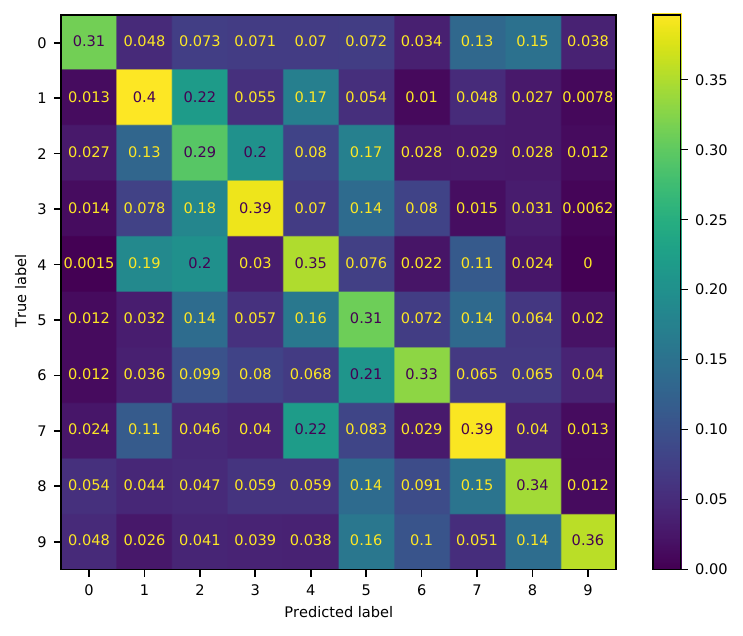}} \quad
\caption{Confusion matrix comparison between our algorithm and humans.}
\label{fig:survey_confusion}
\end{figure*}

\new{In Figure~\ref{fig:survey_confusion}, we compare our model and human performance on the key classification task. The misclassification error and the dispersion result are significantly lower for our algorithm.
Moreover, it can be noticed how the four keys on which humans perform the best match those in the corners of our keypad (i.e., $1$, $3$, $7$, and $9$).}

\section{Additional Experiments}
\label{app:additional_experiments}

\new{To gain further insight into how coverage can affect the attack performance, we grouped the tested users by the coverage strategy:
\begin{compactitem}
\item \texttt{Side}: The non-typing hand rests on the side of the palm and is angled to cover the keys of the PIN pad (40\% of users applied this covering strategy).
\item \texttt{Over}: The non-typing hand is raised completely off the surface, covering the PIN pad both with the entire back of the hand and the fingers (43\% of users applied this covering strategy).
\item \texttt{Top}: The fingers of the non-typing hand rest on the top of the PIN pad, and the back of the hand is used for the coverage (17\% of users applied this covering strategy).
\end{compactitem}
}

\begin{figure*}[htp]
\centering
\subfloat[\emph{Left-corner camera.}\label{fig:left_camera}]
{\includegraphics[width=0.2\textwidth]{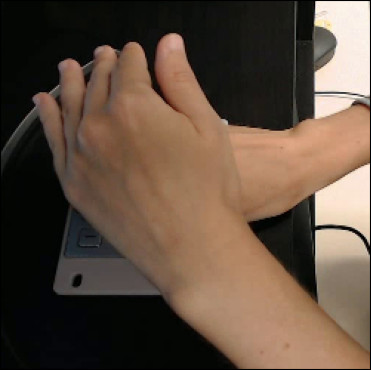}} \quad
\subfloat[\emph{Center camera.}\label{fig:right_camera}]
{\includegraphics[width=0.2\textwidth]{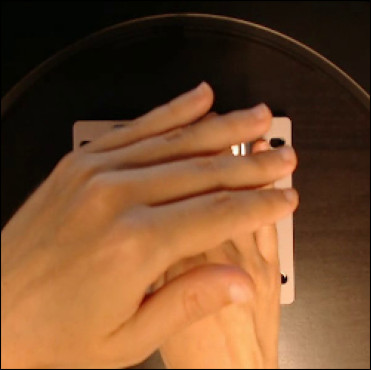}} \quad
\subfloat[\emph{Right-corner camera.}\label{fig:both_cameras}]
{\includegraphics[width=0.2\textwidth]{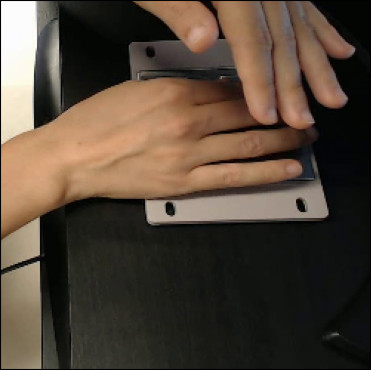}} \quad
\caption{Same video frame recorded by three cameras.}
\label{fig:leftCenterRight}
\end{figure*}

\new{In Table~\ref{tab:coveringStrategies}, we report key and PIN TOP-3 accuracies for our approach.
Clearly, \texttt{Side} covering strategy provides the least protection and should be avoided. At the same time, the \texttt{Over} and \texttt{Top} covering strategies provide much better protection.  Interestingly, we see that with the \texttt{Over} covering strategy, the \textit{Mixed} scenario reaches lower accuracy than the \textit{Single PIN pad} scenario. We postulate this happens as this covering strategy makes it less ``natural'' for the user to type, deceiving the deep learning algorithm. Further attack improvements could be made with datasets having examples of one covering strategy only.
For the \texttt{Top} covering strategy, there were no data for two out of three scenarios (denoted NA in Table~\ref{tab:coveringStrategies}).
}

\begin{table}
\centering
\small
\begin{tabular}{llcc}
\toprule
\multicolumn{1}{c}{\textbf{Covering}} & \multicolumn{1}{c}{\textbf{Scenario}} & \textbf{Key} & \textbf{PIN TOP-3} \\
\multicolumn{1}{c}{\textbf{strategy}} & \multicolumn{1}{c}{\textbf{}} & \textbf{accuracy} & \textbf{accuracy} \\ \midrule
 & Single & 0.64 & 0.30 \\
\texttt{Side} & Independent & 0.42 & 0.12 \\
 & Mixed & 0.77 & 0.53 \\ \hline
 & Single & 0.52 & 0.12 \\
\texttt{Over} & Independent & 0.31 & 0.10 \\
 & Mixed & 0.46 & 0.07 \\ \hline
 & Single & NA & NA \\
\texttt{Top} & Independent & 0.41 & 0.13 \\
 & Mixed & NA & NA \\ \bottomrule
\end{tabular}
\caption{Performance of our attack for different covering strategies in \textit{Single PIN pad}, \textit{PIN pad independent}, and \textit{Mixed} scenarios. \texttt{Top} covering participants were present in the \textit{PIN pad independent} scenario only, as for the others, no data were available (NA).}
\label{tab:coveringStrategies}
\end{table}

\begin{figure}[htp]
\centering
\subfloat[\emph{25\% of PIN pad surface covered (i.e., digits form 1 to 3).}]
{\includegraphics[width=0.2\textwidth]{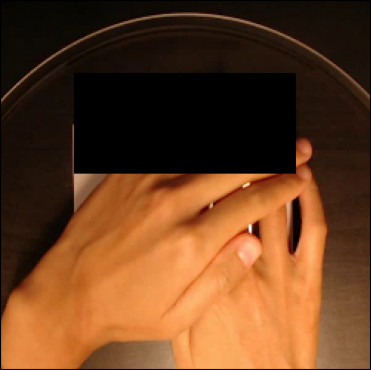}}\quad
\subfloat[\emph{50\% of PIN pad surface covered (i.e., digits form 1 to 6).}]
{\includegraphics[width=0.2\textwidth]{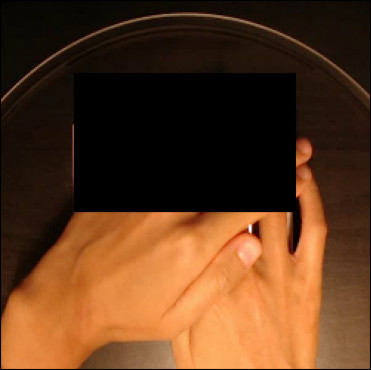}}\\
\subfloat[\emph{75\% of PIN pad surface covered (i.e., digits form 1 to 9).}]
{\includegraphics[width=0.2\textwidth]{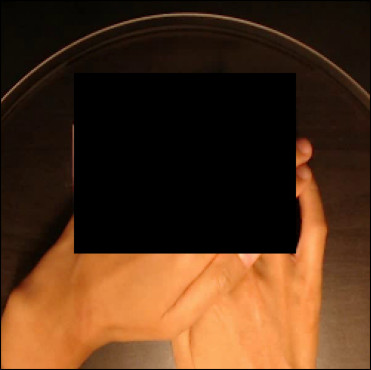}}\quad
\subfloat[\emph{100\% of PIN pad surface covered (i.e., no digit is visible).}]
{\includegraphics[width=0.2\textwidth]{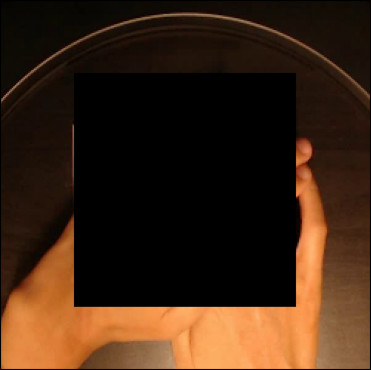}}
\caption{PIN pad shield configurations.}
\label{fig:PINPadShield}
\end{figure}

\new{For the PIN shield countermeasure, we depict various levels of hiding in Figure~\ref{fig:PINPadShield}. There, 25\% denotes that the first row of the PIN pad is covered (simulated with a black patch), 50\% first two rows, 75\% first three rows, and finally, 100\% all four rows of the PIN pad are covered.
Note that we do not include the covering with the other hand into these percentages.}

\new{Table~\ref{tab:generalExperiments} provides results for several additional attack configurations.
First, we performed two experiments simulating a lower camera quality or a larger camera distance from the PIN pad. For this purpose, we reduced the model input resolution from 250 \texttt{x} 250 to 125 \texttt{x} 125 and to 64 \texttt{x} 64. Results show that our model maintains an accuracy higher than 20\%, even when halving the input resolution (i.e., doubling the camera distance). However, this is not to be considered as a physical limitation for our attack since if the attacker places a camera outside the ATM chassis, it is possible to use an optical zoom. Further, many pinhole cameras can record with a resolution up to 1\,080p~\footnote{https://www.dsecctv.com/Prod\_telecamere\_spioncino\_porta\_AHD.htm}, which is higher than the resolution we used to collect our dataset (720p).}

\new{Next, we investigated the accuracy of our attack leveraging different camera positions. In particular, we performed two experiments training and testing our model with the left-corner and the right-corner cameras, respectively. Figure~\ref{fig:leftCenterRight} shows the camera views used in our experiments. The results give a significant difference in performance if the camera is on the right or the left. This is because the participants in our experiment were right-handed, and therefore filming from the right had worse coverage of the PIN pad and typing hand. In contrast, the typing hand and the PIN pad were almost completely covered using shots from the left, significantly reducing the model's performance.
We also evaluated whether using video from all three cameras in training (the experiment ``multi-camera training`` in Table~\ref{tab:generalExperiments}) could improve the accuracy of our model when compared with videos recorded from the center camera only.
The results show a drop in performance, which we attribute to the higher variance in the data provided as input to the model.}

\new{Finally, we report the results of our model without data augmentation and without including the blacklisted users in the training set. In both configurations, the performance of our model drops, showing that reducing the training size is penalized heavily.
Note that even in the worst case of a camera placed on the left corner (i.e., the one with less visibility), \textit{our model still performs better than an average human.}}

\begin{table}
\centering
\small
\begin{tabular}{@{}lcc@{}}
\toprule
\multicolumn{1}{c}{\textbf{Experiment}} & \textbf{Key} & \textbf{PIN TOP-3} \\
\multicolumn{1}{c}{\textbf{}} & \textbf{accuracy} & \textbf{accuracy} \\ \midrule
Input resolution 125 \texttt{x} 125 & 0.55 & 0.23 \\
Input resolution 64 \texttt{x} 64 & 0.47 & 0.15 \\ \midrule
Left-corner camera & 0.46 & 0.10 \\
Right-corner camera & 0.62 & 0.31 \\ \midrule
Multi-camera training & 0.53 & 0.22 \\ \midrule
No data augmentation & 0.44 & 0.11 \\
Blacklisted excluded in training & 0.54 & 0.18 \\ \bottomrule
\end{tabular}
\caption{Additional attack configurations and results in the \textit{Mixed} scenario.}
\label{tab:generalExperiments}
\end{table}

\new{In this paper, we used the feedback sound emitted by the PIN pad as a detection system for the frames containing a keystroke. To evaluate the impact of other frame detection systems, we conducted an experiment varying the frame extraction precision. We simulated the detection error by adding Gaussian noise with mean zero to the ground truth (i.e., the frame position in the video).
In Table~\ref{tab:sigma}, we report the single key and the PIN TOP-3 accuracies for the \textit{Mixed} scenario, simulating five levels of the frame detection error.
Compared to the results obtained using the audio feedback (key accuracy 0.61, 5-digits PIN Top-3 accuracy 0.30), we see that our model works well even with small/medium levels of frame detection error (i.e., less than five frames).
In particular, for a frame error confidence of three (i.e., when the frames are detected through the appearance on the screen of the masked symbols~\cite{cardaioli2020your}), the performance drops only 1\% both for key and TOP-3 PIN accuracies.
Contrarily, when the detection error becomes high (i.e., more than 15 frames), the performance of our model decreases significantly. This happens since the frames considered by the model do not contain information related to the target key, as they are too temporally shifted. Naturally, if the attacker recognizes a situation like this, it would be possible to mitigate the effect of detection error by not using the feedback sound but observing the appearance of ``*'' symbols on the screen.}

\begin{table}
\centering
\small
\begin{tabular}{@{}ccc@{}}
\toprule
\multicolumn{1}{c}{\textbf{Frame error}} & \textbf{Key} & \textbf{PIN TOP-3} \\
\multicolumn{1}{c}{\textbf{confidence (p<0.01)}} & \textbf{accuracy} & \textbf{accuracy} \\ \midrule
$3$ & 0.60 & 0.29 \\
$5$ & 0.59 & 0.26 \\
$10$ & 0.54 & 0.16 \\
$15$ & 0.49 & 0.12 \\
$20$ & 0.12 & 0.06 \\ \bottomrule
\end{tabular}
\caption{Performance of our attack in the \textit{Mixed} scenario assuming different levels of frame detection error.}
\label{tab:sigma}
\end{table}


\end{document}